\documentclass[12pt]{article}

\usepackage{amssymb}

\usepackage[dvips]{graphicx}

\begin{document}

\title
{Derivation of the exact NSVZ $\beta$-function in $N=1$ SQED,
regularized by higher derivatives, by direct summation of Feynman
diagrams}

\author{K.V.Stepanyantz}

\maketitle

\begin{center}

{\em Moscow State University, physical faculty,\\
department of theoretical physics.\\
$119992$, Moscow, Russia}

\end{center}

\begin{abstract}
For $N=1$ supersymmetric quantum electrodynamics, regularized by
higher derivatives, a method for summation of all Feynman diagrams
defining the $\beta$-function is presented. Using this method we
prove that the $\beta$-function is given by an integral of a total
derivative, which can be easily calculated. It is shown that
surviving terms give the exact NSVZ $\beta$-function. The results
are compared with the explicit three-loop calculation.
\end{abstract}


\section{Introduction.}
\hspace{\parindent}

Quantum corrections in supersymmetric theories (and, in
particular, the $\beta$-function) were studied for a long time.
The exact $\beta$-function for $N=1$ supersymmetric theories,

\begin{equation}\label{NSVZ_General_Beta}
\beta(\alpha) = - \frac{\alpha^2\Big[3 C_2 - T(R) + C(R)_i{}^j
\gamma_j{}^i(\alpha)/r\Big)\Big]}{2\pi(1- C_2\alpha/2\pi)},
\end{equation}

\noindent was found in \cite{NSVZ_Instanton}. Certainly, a
$\beta$-function in supersymmetric theories was also calculated
explicitly in the lowest loops. Most calculations were made with
the dimensional reduction \cite{Siegel} in $\overline{MS}$-scheme
\cite{MS}. The NSVZ $\beta$-function agrees with these
calculations in the one- and two-loop approximations. In order to
obtain NSVZ $\beta$-function in higher loops it is necessary to
perform a special redefinition of the coupling constant
\cite{Higher_Loops}. The possibility of such a redefinition is
very nontrivial \cite{Jones}. However, it is well known
\cite{Siegel2} that the dimensional reduction is not
self-consistent. (The dimensional regularization \cite{tHooft}
breaks the supersymmetry and is not convenient for calculations in
supersymmetric theories.) Ways allowing to avoid such problems are
discussed in the literature \cite{Stockinger}.

Other regularizations are also sometimes applied for calculations
in supersymmetric theories. For example, in Ref. \cite{Mas} a
two-loop $\beta$-function of the $N=1$ supersymmetric Yang--Mills
theory was calculated with the differential renormalization
\cite{DiffR}. Some calculations were made with the higher
covariant derivative regularization, proposed in \cite{Slavnov},
which was generalized to the supersymmetric case in Ref.
\cite{Krivoshchekov} (another variant was proposed in Ref.
\cite{West_Paper}). Usually integrals arising with the higher
covariant derivative regularization can not be calculated
analytically in higher loops. That is why this regularization was
applied for explicit calculations rather rarely. In particular,
the first calculation of quantum corrections for the
(non-supersymmetric) Yang--Mills theory was made in Ref.
\cite{Ruiz}. Taking into account corrections, made in subsequent
papers \cite{Asorey}, the result for the $\beta$-function appeared
to be the same as the well-known result obtained with the
dimensional regularization \cite{Politzer}. In principle, it is
possible to prove that in the one-loop approximation calculations
with the higher covariant derivative regularization always agree
with the results of calculations with the dimensional
regularization \cite{PhysLett}. Some calculations in the one- and
two-loop approximations were made for various theories
\cite{Rosten_Calc,Rosten_Int} with a variant of the higher
covariant derivative regularization, proposed in
\cite{Rosten_Reg}. The structure of the corresponding integrals
was discussed in Ref. \cite{Rosten_Int}.

The three-loop $\beta$-function for the $N=1$ SQED, regularized by
higher derivatives, was calculated in \cite{3LoopHEP}. This
calculation shows that the integrals defining the $\beta$-function
are integrals of total derivatives. A two-loop calculation made
with the dimensional reduction in the $N=1$ SQED and revealing a
similar feature was presented in Ref. \cite{Smilga}, where the
factorization of integrands into total derivatives is explained in
all loops using a special technique, based on the covariant
Feynman rules in the background field method \cite{Grisaru}. This
factorization allows to calculate one of the loop integrals
analytically. As a consequence, (in the $N=1$ SQED) integrals
defining the $\beta$-function are reduced to integrals defining
the anomalous dimension, producing NSVZ $\beta$-function. In
particular, with the higher derivative regularization it is not
necessary to perform a redefinition of the coupling constant. The
factorization of integrands into total derivatives seems to be a
general feature of supersymmetric theories. In the two-loop
approximation it was verified for a general renormalizable $N=1$
supersymmetric theory, regularized by higher covariant
derivatives, in \cite{PhysLettSUSY}.

Using a usual supergraph technique \cite{West,Buchbinder} an
attempt to prove the factorization of integrands into total
derivatives was made in Ref. \cite{SD}, where a solution of the
Ward identity was substituted into the Schwinger--Dyson equations.
This allows to present a $\beta$-function as a sum of two
contributions: The first one (calculated in Ref. \cite{SD} exactly
to all orders) is given by an integral of a total derivative and
is expressed in terms of the two-point function of the matter
superfields. This contribution gives the exact NSVZ
$\beta$-function. The second contribution is essential starting
from the three-loop approximation and can be expressed in terms of
a transversal part of a certain Green function. Explicit
three-loop calculations \cite{3LoopHEP} show that the second
contribution to the $\beta$-function is also given by an integral
of a total derivative and is equal to 0. It was conjectured
\cite{SD} that this takes place in all loops. A partial four-loop
verification of this statement were made in Ref. \cite{Pimenov}.
The factorization of the additional contribution into an integral
of a total derivative was qualitatively explained in
\cite{FactorizationHEP}, using a technique proposed in Ref.
\cite{Identity}. In this paper we formulate these arguments in a
rigorous form and prove that all integrals in the $N=1$ SQED
defining the $\beta$-function are integrals of total derivatives.
Taking these integrals we obtain the exact NSVZ $\beta$-function
without a redefinition of the coupling constant. The results are
compared with the explicit three-loop calculation, made in
\cite{3LoopHEP}.

The paper is organized as follows:

In Sec. \ref{Section_SUSY_QED} we recall basic information about
the $N=1$ supersymmetric electrodynamics and its regularization by
higher derivatives. In Sec. \ref{Section_Generating_Functional} we
perform the integration over the matter superfields in the
generating functional and introduce some notation. Then it is
possible to construct formal expressions for a $\beta$-function
and an anomalous dimension, encoding sums of Feynman diagrams.
This is made in Sec. \ref{Section_Formal_Expressions}. In Sec.
\ref{Section_Tricks} we describe some tricks allowing to simplify
the calculation of the Feynman diagrams. In the massless case
these diagrams are calculated (exactly to all orders) in Sec.
\ref{Section_Calculation}. In this section the sum of diagrams is
reduced to integrals of total derivatives. From these integrals of
total derivatives the exact NSVZ $\beta$-function is derived
exactly to all orders in Sec. \ref{Section_NSVZ}. A similar
investigation for the Pauli--Villars contributions is made in Sec.
\ref{Pauli_Villars}. The result is also given by integrals of
total derivatives. The obtained expressions are verified by the
explicit three-loop calculation in Sec. \ref{Section_Three_Loop}.
The results are summarized in the Conclusion. Some technical
details are presented in the Appendixes.

\section{$N=1$ supersymmetric electrodynamics and its regularization
by higher derivatives} \label{Section_SUSY_QED}
\hspace{\parindent}

The massless $N=1$ supersymmetric electrodynamics is described by
the following action:\footnote{In our
notation $\eta_{\mu\nu}=\mbox{diag}(1,-1,-1,-1)$;\ \ $\theta^a
\equiv \theta_b C^{ba}$; $\theta_a$ and $\bar\theta_a$ denote the
right and left components of $\theta$, respectively.}

\begin{equation}\label{SQED_Action}
S = \frac{1}{4 e^2} \mbox{Re}\int d^4x\,d^2\theta\,W_a C^{ab} W_b
+ \frac{1}{4}\int d^4x\, d^4\theta\, \Big(\phi^* e^{2V}\phi
+\widetilde\phi^* e^{-2V}\widetilde\phi\Big).
\end{equation}

\noindent Here $\phi$ and $\widetilde\phi$ are chiral matter
superfields, and $V$ is a real scalar superfield, which contains
the gauge field $A_\mu$ as a component. The superfield $W_a$ is a
supersymmetric analog of the gauge field strength. In the Abelian
case it is defined by

\begin{equation}
W_a = \frac{1}{4} \bar D^2 D_a V.
\end{equation}

\noindent ($D_a$ and $\bar D_a$ are the right and left
supersymmetric covariant derivatives, respectively.)

In order to regularize the theory we add to the action a term with
higher derivatives. Then the action can be written as

\begin{equation}\label{Regularized_SQED_Action}
S_{\mbox{\scriptsize reg}} = \frac{1}{4 e^2} \mbox{Re}\int
d^4x\,d^2\theta\,W_a C^{ab}
R\Big(\frac{\partial^{2}}{\Lambda^{2}}\Big) W_b + \frac{1}{4}\int
d^4x\, d^4\theta\, \Big(\phi^* e^{2V}\phi +\widetilde\phi^*
e^{-2V}\widetilde\phi\Big),
\end{equation}

\noindent where the function $R$ satisfies the following
conditions:

\begin{equation}
R(0) = 1;\qquad R(\infty) = \infty.
\end{equation}

\noindent For example, it is possible to choose

\begin{equation}
R\Big(\frac{\partial^{2}}{\Lambda^{2}}\Big) = 1 +
\frac{\partial^{2n}}{\Lambda^{2n}}.
\end{equation}

\noindent In the Abelian case the superfield $W_a$ is gauge
invariant, so that action (\ref{Regularized_SQED_Action}) is also
gauge invariant.

Quantization of model (\ref{Regularized_SQED_Action}) can be made
by the standard way. For this purpose it is convenient to use the
supergraph technique, described, for example, in textbooks
\cite{West,Buchbinder}, and to fix the gauge invariance by adding
the following terms:

\begin{equation}\label{Gauge_Fixing}
S_{\mbox{\scriptsize gf}} = - \frac{1}{64 e^2}\int
d^4x\,d^4\theta\, \Big(V D^2 \bar D^2
R\Big(\frac{\partial^{2}}{\Lambda^{2}}\Big) V + V \bar D^2 D^2
R\Big(\frac{\partial^{2}}{\Lambda^{2}}\Big) V\Big).
\end{equation}

\noindent After adding such terms, a kinetic term for the
superfield $V$ will have the simplest form

\begin{equation}
S_{\mbox{\scriptsize reg}} + S_{\mbox{\scriptsize gf}} =
\frac{1}{4 e^2}\int d^4x\,d^4\theta\, V\partial^2
R\Big(\frac{\partial^{2}}{\Lambda^{2}}\Big) V.
\end{equation}

\noindent In the Abelian case, considered here, diagrams with
ghost loops are absent.

Adding the higher derivative term does not remove divergences in
the one-loop diagrams. In order to regularize them, it is
necessary to insert Pauli-Villars determinants into the generating
functional \cite{Slavnov_Book}. Therefore, the generating
functional can be written as

\begin{equation}\label{Modified_Z}
Z = \int DV\,D\phi\,D\widetilde \phi\, \prod\limits_I \Big(\det
PV(V,M_I)\Big)^{c_I} \exp\Big(i(S_{\mbox{\scriptsize
reg}}+S_{\mbox{\scriptsize gf}}+S_{\mbox{\scriptsize
Source}})\Big),
\end{equation}

\noindent where

\begin{equation}
S_{\mbox{\scriptsize Sourse}} = \int d^4x\,d^4\theta\,V J +
\Big(\int d^4x\,d^2\theta\,(\phi\,j +\widetilde\phi\,\widetilde j)
+\mbox{h.c.}\Big).
\end{equation}

\noindent (It is necessary to substitute $e$ in
$S_{\mbox{\scriptsize reg}}$ and $S_{\mbox{\scriptsize gf}}$ by
the bare coupling constant $e_0$.) The Pauli-Villars determinants
are given by

\begin{equation}\label{PV_Determinants}
\Big(\det PV(V,M)\Big)^{-1} = \int D\Phi\,D\widetilde \Phi\, e^{i
S_{\mbox{\scriptsize PV}}},
\end{equation}

\noindent where\footnote{Note that the Pauli--Villars action
differs from the one used in \cite{3LoopHEP} because here the
ratio of the coefficients in the kinetic term and in the mass term
does not contain the factor $Z$. Using terminology of Ref.
\cite{Arkani}, one can say that here we will calculate the
canonical coupling $\alpha_c$, while in Ref. \cite{3LoopHEP} the
holomorphic coupling $\alpha_h$ have been calculated. Certainly,
after the renormalization the effective action does not depend on
the definitions. However, the definitions used here are much more
convenient.}

\begin{equation}
S_{\mbox{\scriptsize PV}}\equiv \frac{1}{4} \int d^4x\,d^4\theta\,
\Big(\Phi^* e^{2V}\Phi + \widetilde\Phi^* e^{-2V}\widetilde\Phi
\Big) + \Big(\frac{1}{2}\int d^4x\,d^2\theta\, M \widetilde\Phi
\Phi + \mbox{h.c.}\Big).
\end{equation}

\noindent The coefficients $c_I$ satisfy conditions

\begin{equation}
\sum\limits_I c_I = 1;\qquad \sum\limits_I c_I M_I^2 = 0.
\end{equation}

\noindent Below we will assume, that $M_I = a_I\Lambda$, where
$a_I$ are constants. Insertion of the Pauli-Villars determinants
allows to cancel remaining divergences in all one-loop diagrams.

The generating functional for connected Green functions and the
effective action are defined by the standard way.

\section{Generating functional}
\hspace{\parindent}\label{Section_Generating_Functional}

In order to derive an exact $\beta$-function we perform explicit
summation of the corresponding Feynman diagrams. However, for the
rigorous proof it is desirable to obtain some formal expressions
encoding sums of these diagrams. For this purpose we first perform
the integration over the matter superfields in generating
functional (\ref{Modified_Z}). This can be made, because the
corresponding integral is Gaussian. It is easy to see that the
result is

\begin{eqnarray}\label{Z_Without_Phi}
&& Z = \int DV\,\prod\limits_I \Big(\det PV(V,M_I)\Big)^{c_I}
\det(*)\det(\widetilde *)
\nonumber\\
&& \times \exp\Bigg\{ i\int d^8x\,\Big(\frac{1}{4e_0^2} V\partial^2
R(\partial^2/\Lambda^2) V - j \frac{D^2}{4\partial^2} * \frac{\bar
D^2}{4\partial^2} j^* - \widetilde j \frac{D^2}{4\partial^2}
\widetilde * \frac{\bar D^2}{4\partial^2} \widetilde j^* \Big)
\Bigg\},\qquad
\end{eqnarray}

\noindent where

\begin{equation}
* \equiv \frac{1}{1-(e^{2V}-1)\bar D^2 D^2/16\partial^2},
\qquad \widetilde * = \frac{1}{1-(e^{-2V}-1)\bar D^2
D^2/16\partial^2}
\end{equation}

\noindent encode chains of propagators, connecting vertexes with
the quantum gauge field.\footnote{This is a rigorous definition.
In Ref. \cite{FactorizationHEP} $*$ was defined only qualitatively
using Feynman rules. That is why some equations below differ from
the corresponding equations in Ref. \cite{FactorizationHEP}.} In
order to obtain generating functional (\ref{Z_Without_Phi}), we
note that a solution of the motion equation for the chiral
superfield $\phi$

\begin{equation}
8 j^* = D^2 (e^{2V} \phi) = D^2\phi + D^2 \Big((e^{2V}-1)\phi\Big)
\end{equation}

\noindent can be written as

\begin{equation}
\phi = - \Big(1-\frac{\bar D^2 D^2}{16\partial^2}
(e^{2V}-1)\Big)^{-1} \frac{\bar D^2}{2\partial^2} j^* = \frac{\bar
D^2 D^2}{16\partial^2} * \frac{\bar D^2}{2\partial^2} j^*.
\end{equation}

Similar to derivation of Eq. (\ref{Z_Without_Phi}) it is possible
to perform integration over the Pauli--Villars fields. Introducing
sources for the Pauli--Villars fields in expression
(\ref{PV_Determinants}), we can write the Pauli--Villars
determinants in a similar form:

\begin{equation}\label{Expression_For_PV_Determinants}
\det PV(V,M,\mbox{\boldmath$j$}_{\mbox{\scriptsize pv}}) =
\Big(\det(\star)\Big)^{1/2}\exp\Big\{-\frac{i}{2}\int d^8x\,
\mbox{\boldmath$j$}_{\mbox{\scriptsize pv}}^T A P \star A
\mbox{\boldmath$j$}_{\mbox{\scriptsize pv}} \Big\},
\end{equation}

\noindent where we use the following notation:

\begin{equation}
A \equiv \frac{1}{4\partial^2}\left(
\begin{array}{cccc}
D^2 & 0 & 0 & 0\\
0 & \bar D^2 & 0 & 0\\
0 & 0 & D^2 & 0\\
0 & 0 & 0 & \bar D^2
\end{array}
\right);\qquad \mbox{\boldmath$j$} = \left(
\begin{array}{c}
j\\
j^*\\
\widetilde j\\
\widetilde j^*
\end{array}
\right).
\end{equation}

\noindent $\star$ is defined by

\begin{equation}\label{Star_Definition}
\star \equiv \frac{1}{1-I_0} \qquad \mbox{with} \qquad I_0 = {\cal
V} P,
\end{equation}

\noindent where

\begin{equation}\label{Nu_Definition}
{\cal V} = \left(
\begin{array}{cccc}
0 & (e^{2V}-1) & 0 & 0\\
(e^{2V}-1) & 0 & 0 & 0\\
0 & 0 & 0 & (e^{-2V}-1)\\
0 & 0 & (e^{-2V}-1) & 0
\end{array}
\right)
\end{equation}

\noindent is a vertex contribution, and

\begin{equation}
P = \left(
\begin{array}{cccc}
0 & {\displaystyle \frac{\bar D^2 D^2}{16(\partial^2 + M^2)}}
& {\displaystyle \frac{M \bar D^2}{4(\partial^2 + M^2)}} & 0\vphantom{\Bigg(}\\
{\displaystyle \frac{D^2 \bar D^2}{16(\partial^2 + M^2)}} & 0 & 0
& {\displaystyle \frac{M D^2}{4(\partial^2 + M^2)}}\vphantom{\Bigg(}\\
{\displaystyle \frac{M \bar D^2}{4(\partial^2 + M^2)}}
& 0 & 0 & {\displaystyle \frac{\bar D^2 D^2}{16(\partial^2 + M^2)}} \vphantom{\Bigg(}\\
0 & {\displaystyle \frac{M D^2}{4(\partial^2 + M^2)}} &
{\displaystyle \frac{D^2 \bar D^2}{16(\partial^2 + M^2)}} &
0\vphantom{\Bigg(}
\end{array}
\right)
\end{equation}

\noindent corresponds to propagators.

The contribution of $\phi$ and $\widetilde\phi$ can be also
written in a form similar to
(\ref{Expression_For_PV_Determinants}):

\begin{eqnarray}
&& Z = \int DV\,\prod\limits_I \Big(\det PV(V,M_I)\Big)^{c_I}
\Big(\det(\star)\Big)^{1/2}
\nonumber\\
&& \times \exp\Bigg\{ i\int d^8x\,\frac{1}{4e_0^2} V\partial^2
R(\partial^2/\Lambda^2) V +\frac{i}{2}\int d^8x\,
\mbox{\boldmath$j$}^T A P \star A \mbox{\boldmath$j$}
\Bigg\},\qquad
\end{eqnarray}

\noindent where $M$ inside $\star$ should be set to 0.

It is also convenient to define

\begin{eqnarray}
&& (I_1)_a \equiv [I_0,\theta_a];\qquad (\bar I_1)_a \equiv
[I_0,\bar\theta_a];\qquad (I_2) \equiv
\{[I_0,\theta_a],\theta^a\};\vphantom{\Big(}\nonumber\\
&& (I_2)_{ab} \equiv \{[I_0,\bar\theta_a],\theta_b\};\qquad (\bar
I_3)_b \equiv
[\{[I_0,\theta_a],\theta^a\},\bar\theta_b].\vphantom{\Big(}
\end{eqnarray}

\noindent If, for example\footnote{This expression can be used in
the massless case, if only the field $\phi$ is present. In the
general case $I_0$ is defined by Eq. (\ref{Star_Definition}).},
$I_0 = (e^{2V}-1) \bar D^2 D^2/16\partial^2$, then

\begin{eqnarray}
&& (I_1)_a = (e^{2V}-1)\frac{\bar D^2 D_a}{8
\partial^2};\qquad (\bar I_1)_a = (e^{2V}-1)\frac{\bar
D_a D^2}{8 \partial^2};\qquad (I_2)
= (e^{2V}-1)\frac{\bar D^2}{4\partial^2};\nonumber\\
&& (I_2)_{ab} = (e^{2V}-1)\frac{\bar D_a D_b}{4\partial^2};\qquad
(\bar I_3)_b = (e^{2V}-1)\frac{\bar D_b}{2\partial^2}.
\end{eqnarray}

Let us also describe some properties of $\star$, which will be
useful below:

It is easy to see that

\begin{equation}
\star I_0 \star = - \star + \star^2,
\end{equation}

\noindent and the following expansions take place:

\begin{equation}
\star = 1+ \sum\limits_{n=1}^\infty (I_0)^n;\qquad \ln \star =
\sum\limits_{n=1}^\infty \frac{1}{n} (I_0)^n;\qquad \star^2 =
\sum\limits_{n=0}^\infty (n+1) (I_0)^n,\quad \mbox{etc}.
\end{equation}

\noindent Therefore,

\begin{eqnarray}\label{Star_Coefficients}
&& (\ln \star)_n = \frac{1}{n}(\star)_n;\qquad (\star^2)_n = (n+1)
(\star)_n;\qquad (\star^3)_n = \frac{(n+1)(n+2)}{2} (\star)_n; \nonumber\\
&& \qquad\qquad\qquad\quad (\star^4)_n = \frac{(n+1)(n+2)(n+3)}{6}
(\star)_n,
\end{eqnarray}

\noindent where the subscript $n$ denotes the $n$-th term.

\section{Formal calculation of renormgroup function}
\hspace{\parindent}\label{Section_Formal_Expressions}

Expression (\ref{Z_Without_Phi}) allows to obtain a simple formal
expression for an anomalous dimension. A Green function of the
matter superfield is given by

\begin{equation}
\frac{\delta^2\Gamma}{\delta\phi_{x} \delta\phi_y^*} = \frac{\bar
D_x^2 D_x^2}{16}G(\partial_x^2)\delta^8_{xy}.
\end{equation}

\noindent The corresponding inverse function, which by definition
satisfies

\begin{equation}
\int d^8y\,\frac{\delta^2\Gamma}{\delta\phi_x \delta\phi_y^*}
\frac{\bar D_y^2}{8\partial^2}
\Bigg(\frac{\delta^2\Gamma}{\delta\phi_z \delta\phi_y^*}
\Bigg)^{-1} = - \frac{1}{2}\bar D_x^2 \delta^8_{xz},
\end{equation}

\noindent is

\begin{equation}
\Bigg(\frac{\delta^2\Gamma}{\delta\phi_y^*
\delta\phi_x}\Bigg)^{-1} = - \frac{\bar D_x^2
D_x^2}{4\partial^2}G^{-1}(\partial^2)\delta^8_{xy} =
-\frac{\delta^2 W}{\delta j_y^* \delta j_x} = \Big\langle
\frac{\bar D_x^2 D_x^2}{8\partial^2} * \frac{\bar D_x^2
D_x^2}{8\partial^2} \delta^8_{xy}\Big\rangle.
\end{equation}

\noindent (In order to derive the last equality, we have
differentiated generating functional (\ref{Z_Without_Phi}) with
respect to the sources.) The angular brackets denote the
functional integration over the gauge superfield $V$. (The factors
$\det(*)$ and $\det(\widetilde *)$ should be certainly included.)
From this equation we obtain

\begin{equation}\label{Formal_Anomalous_Dimension}
\bar D_x^2 D_x^2 G^{-1}(\partial^2) \delta^8_{xy} = \Big\langle
* \bar D_x^2 D_x^2 \delta^8_{xy}\Big\rangle.
\end{equation}

\noindent This can be verified by applying to both sides the
operator $- \bar D_x^2 D_x^2/16\partial_y^2$. We will use
expression (\ref{Formal_Anomalous_Dimension}) later.

It is also possible to construct an expression for a two-point
Green function of the gauge superfield. For this purpose we use
the equation

\begin{eqnarray}\label{Gamma_2_V}
\Gamma^{(2)}_{\bf V} = \frac{1}{4e_0^2} \int d^8x\, {\bf
V}\partial^2 R {\bf V} + \frac{1}{2} \int d^8x\,d^8y\,{\bf V}_x
{\bf V}_y \Big\langle i \frac{\delta S_I}{\delta V_x} \frac{\delta
S_I}{\delta V_y} + \frac{\delta^2 S_I}{\delta V_x \delta V_y}
\Big\rangle_{\mbox{\scriptsize 1PI}},
\end{eqnarray}

\noindent which is derived in Appendix \ref{Appendix_Gamma_2_V}.
Here the symbol 1PI means that in this expression it is necessary
to keep only one-particle irreducible graphs, and $S_I$ denotes
the interaction. For convenience we denoted the argument of the
effective action by the bold letter ${\bf V}$. (This equation can
be also easily obtained using the background field method, which
is not used in this paper.) Here

\begin{equation}
S_I = \frac{1}{4} \int d^8x\,\Big(\phi^* (e^{2V}-1)\phi +
\widetilde\phi^* (e^{-2V} -1) \tilde\phi \Big).
\end{equation}

\noindent We substitute this expression into Eq.
(\ref{Gamma_2_V}), taking into account the identity

\begin{equation}
\langle f(\phi,\phi^*) \rangle = \frac{1}{Z}
f\Big(\frac{1}{i}\frac{\delta}{\delta
j},\frac{1}{i}\frac{\delta}{\delta j^*}\Big) Z,
\end{equation}

\noindent where the generating functional $Z$ is given by Eq.
(\ref{Z_Without_Phi}). After simple calculations we
obtain\footnote{In order to find the contributions of $\phi$- and
$\widetilde\phi$-loops it is necessary to set $M=0$.}

\begin{equation}
\Gamma^{(2)}_{\bf V} = S_{\bf V}^{(2)} + S_{\mbox{\scriptsize gf}}
+\Big\langle - \frac{i}{2} \Big(\mbox{Tr}({\bf V} Q J_0
\star)\Big)^2 - i \mbox{Tr}({\bf V} Q J_0 \star {\bf V} Q J_0
\star) - i \mbox{Tr}({\bf V}^2 J_0 \star)\Big\rangle + (PV),
\end{equation}

\noindent where

\begin{equation}
\mbox{Tr}\, A = \mbox{tr}\int d^8x\,A_{xx},
\end{equation}

\noindent and $\mbox{tr}$ denotes a usual matrix trace (if it is
needed). $(PV)$ denotes contributions of the Pauli--Villars
fields,

\begin{equation}\label{Q_And_J0_Definition}
Q = \left(
\begin{array}{cccc}
1 & 0 & 0 & 0\\
0 & 1 & 0 & 0\\
0 & 0 & -1 & 0\\
0 & 0 & 0 & -1
\end{array}
\right);\qquad J_0 = \left(
\begin{array}{cccc}
0 & e^{2V} & 0 & 0\\
e^{2V} & 0 & 0 & 0\\
0 & 0 & 0 & e^{-2V}\\
0 & 0 & e^{-2V} & 0
\end{array}
\right) P.
\end{equation}

Due to the supersymmetric Ward identity the two-point Green
function of the gauge superfield can be presented in the following
form:

\begin{equation}\label{D_Definition}
\Gamma^{(2)}_{\bf V} - S_{\mbox{\scriptsize gf}} = -
\frac{1}{16\pi} \int \frac{d^4p}{(2\pi)^4}\,d^4\theta\,{\bf
V}(\theta,-p)\,\partial^2\Pi_{1/2} {\bf V}(\theta,p)\,
d^{-1}(\alpha,\lambda,\mu/p),
\end{equation}

\noindent where $\alpha$ is a renormalized coupling constant, and

\begin{equation}
\partial^2\Pi_{1/2}  = -\frac{1}{8} D^a \bar D^2 D_a
\end{equation}

\noindent is a supersymmetric transversal projector. We will
calculate the expression

\begin{equation}\label{We_Calculate}
\frac{d}{d\ln \Lambda}\,
\Big(d^{-1}(\alpha_0,\Lambda/p)-\alpha_0^{-1}\Big)\Big|_{p=0} = -
\frac{d\alpha_0^{-1}}{d\ln\Lambda} =
\frac{\beta(\alpha_0)}{\alpha_0^2}.
\end{equation}

\noindent (Here $\Lambda$ and $\alpha$ are considered as
independent variables.) From this equation it is evident that this
expression is well defined. (Later we will demonstrate this in the
three-loop approximation explicitly.) Note that here we implicitly
use the higher derivative regularization, because it allows to
perform differentiation with respect to $\ln\Lambda$ and set the
external momentum $p$ to 0.

\section{Some useful tricks and summation of subdiagrams}
\hspace{\parindent}\label{Section_Tricks}

In order to calculate expression (\ref{We_Calculate}) we consider

\begin{equation}
\frac{d}{d\ln\Lambda}\Big(\Gamma^{(2)}_{\bf V} - S -
S_{\mbox{\scriptsize gf}} \Big)\Big|_{p=0}.
\end{equation}

\noindent Making calculations in the limit $p\to 0$, where $p$ is
the external momentum, is possible, because the corresponding
integrals are well defined in this limit. The higher derivative
regularization and the differentiation with respect to
$\ln\Lambda$ ensure that there are no IR divergences. This agrees
with the results of Ref. \cite{Fargnoli} that the IR region does
not affect to the $\beta$-function. In order to obtain a
transversal part of the two-loop Green function of the gauge
superfield by the simplest way, we make the substitution

\begin{equation}\label{Substitution}
{\bf V}(x,\theta) \to \bar\theta^a\bar\theta_a
\theta^b\theta_b\equiv \theta^4,
\end{equation}

\noindent so that

\begin{equation}
\int d^4\theta\,{\bf V}(x,\theta)\partial^2\Pi_{1/2} {\bf
V}(x,\theta) \to -8.
\end{equation}

\noindent  (This is possible, because in the limit $p\to 0$ the
gauge superfield ${\bf V}$ does not depend on the coordinates
$x^\mu$.) In the momentum representation

\begin{equation}
{\bf V}(p,\theta) = \int d^4x\,{\bf V}(x,\theta) e^{-ip_\alpha
x^\alpha} \to (2\pi)^4 \delta^{4}(p) \theta^4.
\end{equation}

\noindent Thus, after substitution (\ref{Substitution}) we obtain

\begin{eqnarray}\label{We_Calculate2}
&& (2\pi)^3\delta^{4}(p) \frac{d}{d\ln \Lambda}\,
\Big(d^{-1}(\alpha_0,\Lambda/p)-\alpha_0^{-1}\Big)\Big|_{p=0} =
(2\pi)^3\delta^{4}(p)
\frac{\beta(\alpha_0)}{\alpha_0^2}\Bigg|_{p=0}
\nonumber\\
&& = \frac{d}{d\ln\Lambda}\Big(\Gamma^{(2)}_{\bf V} - S -
S_{\mbox{\scriptsize gf}} \Big)\Big|_{p=0, {\bf
V}(x,\theta)=\theta^4}.
\end{eqnarray}

We will try to reduce the sum of Feynman diagrams for the
considered theory to integrals of total derivatives. In the
coordinate representation such an integral can be written as

\begin{equation}\label{Commutator}
\mbox{Tr} \Big([x^\mu, \mbox{Something}]\Big) = 0.
\end{equation}

In order to find a $\beta$-function one should consider the
massless theory. In the massless limit the fields $\phi$ and
$\widetilde\phi$ decouple. The Pauli--Villars contributions (for
which this is not true) will be considered later. First, we will
find a contribution of the field $\phi$ to the $\beta$-function.
The contribution of the field $\widetilde\phi$ can be found
similarly. We will take it into account in the end.

In order to extract commutators (\ref{Commutator}), we consider
diagrams containing a vertex to that only one external line (and
no internal lines) is attached. We can add such a diagram to a
diagram, in which the external line is shifted to the nearest
vertex. Let us formulate this rigorously. In the massless case

\begin{equation}
J_0 \to e^{2V} \frac{\bar D^2 D^2}{16\partial^2};\qquad I_0 \to
(e^{2V}-1) \frac{\bar D^2 D^2}{16\partial^2};\qquad P \to
\frac{\bar D^2 D^2}{16\partial^2};\qquad {\cal V}\to (e^{2V}-1),
\end{equation}

\noindent so that $J_0 = I_0 + P$. As a consequence,

\begin{equation}\label{Sum_Of_Subdiagrams}
* {\bf V} J_0 = \frac{1}{1-I_0} {\bf V} (I_0 + P) = {\bf V} P +
\frac{1}{1-I_0} \Big(I_0 {\bf V} P + {\bf V} I_0\Big) = {\bf V} P
+ * {\cal V}\Big(P {\bf V} P + {\bf V} P\Big).
\end{equation}

\noindent The expression ${\cal V}(P {\bf V} P + {\bf V} P)$
corresponds to a sum of subdiagrams presented below. Making a
substitution ${\bf V} \to \bar\theta^a\bar\theta_a
\theta^b\theta_b$ we obtain

\vspace*{5mm}

\begin{picture}(0,0)
\put(-0.5,-1.5){\includegraphics[scale=0.4]{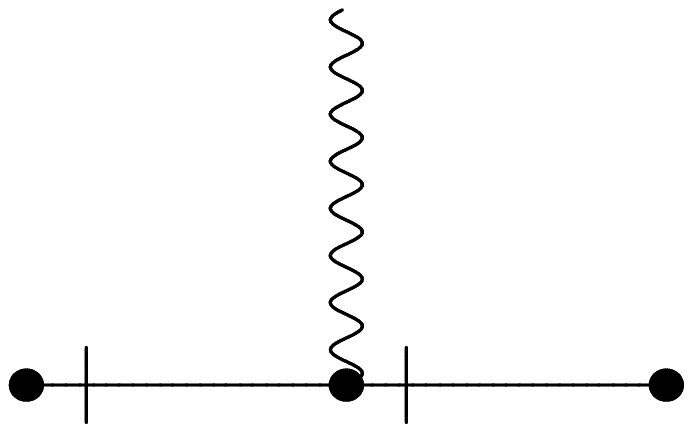}}

\put(3.1,-0.5){$+$}

\put(3.5,-1.5){\includegraphics[scale=0.4]{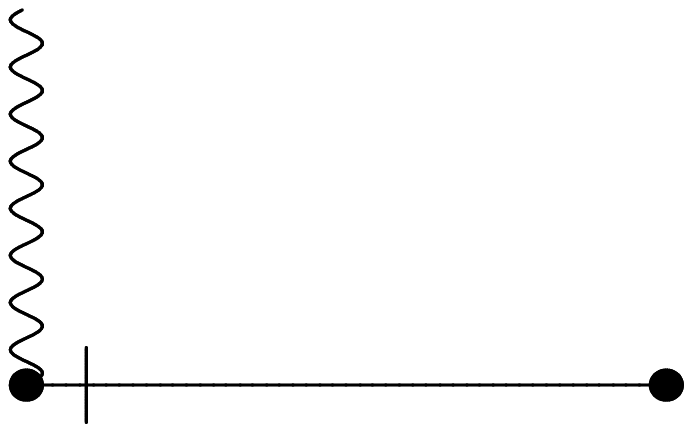}}

\put(7.2,-0.5){$=$}

\end{picture}

\vspace{8mm}
\begin{equation}\label{Subdiagrams_Massless}
=(e^{2V}-1)\Big( -\theta^a \theta_a \bar\theta^b \frac{\bar D_b
D^2}{4\partial^2} + \theta^a \theta_a \frac{D^2}{4\partial^2} + i
\bar\theta^b (\gamma^\mu)_b{}^a \theta_a \frac{\bar D^2 D^2
\partial_\mu}{8\partial^4} - i\theta^a
(\gamma^\mu)_a{}^b \frac{\bar D_b D^2
\partial_\mu}{4\partial^4} + \frac{\bar D^2
D^2}{16\partial^4}\Big).
\end{equation}

\vspace*{5mm}

Only the first and the third terms give nontrivial contributions
to the two-point function of the gauge superfield, because they
contain $\bar\theta$. Really, finally it is necessary to obtain
$$
\int d^4\theta\,\bar\theta^a \bar\theta_a \theta^b \theta_b,
$$
while calculating a $\theta$-part of a graph can not increase
degrees of $\theta$ or $\bar\theta$. Therefore, we should have
$\bar\theta^a \bar\theta_a$ from the beginning.

\section{Reducing the sum of diagrams to integrals of total derivatives}
\label{Section_Calculation}

\subsection{One-loop approximation}\hspace{\parindent}

For the general renormalizable $N=1$ supersymmetric Yang-Mills
theory, regularized by higher derivatives, the one-loop
$\beta$-function was calculated in \cite{PhysLettSUSY}. The result
is given by an integral of a total derivative and agrees with the
exact NSVZ $\beta$-function. Therefore, below we can make
calculations starting from the two-loop approximation.

\subsection{External ${\bf V}$-lines are attached to
different loops of the matter superfields} \hspace{\parindent}

Let us try to find a sum of Feynman diagrams exactly to all orders
of the perturbation theory. We will start with diagrams in that
the external lines are attached to different loops of matter
superfields. Let us consider a loop of matter superfields with $n$
vertexes. This loop is proportional to

\begin{eqnarray}
&& \mbox{Tr}\Big\langle i\bar\theta^c (\gamma^\nu)_c{}^d \theta_d
(e^{2V}-1)\frac{\bar D^2 D^2
\partial_\nu}{8\partial^4} * - \theta^c\theta_c \bar\theta^d
(e^{2V}-1)\frac{\bar D_d D^2}{4\partial^2} *\Big\rangle_n  =\nonumber\\
&& = \frac{1}{n}\mbox{Tr}\Big\langle i\bar\theta^c
(\gamma^\nu)_c{}^d \theta_d (e^{2V}-1) \frac{\bar D^2 D^2
\partial_\nu}{8\partial^2} *^2 - \theta^c\theta_c \bar\theta^d
(e^{2V}-1)\frac{\bar D_d D^2}{4\partial^2} *^2
\Big\rangle_n.\qquad
\end{eqnarray}

\noindent ($*^2$ contains $n-1$ vertexes, and one vertex
corresponds to explicitly written $(e^{2V}-1)$.) After simple
algebraic transformations this expression can be written as

\begin{eqnarray}
&& \frac{1}{n}\mbox{Tr}\Big\langle - \theta^c\theta_c \bar\theta^d
* (e^{2V}-1)\frac{\bar D_d D^2}{4\partial^2} *
- \bar\theta^d \theta^c * (e^{2V}-1)\frac{\bar D^2
D_c}{4\partial^2}
* (e^{2V}-1)\frac{\bar D_d D^2}{4\partial^2} * +\nonumber\\
&& + i\bar\theta^c (\gamma^\nu)_c{}^d \theta_d *
(e^{2V}-1)\frac{\bar D^2 D^2
\partial_\nu}{8\partial^4} *
+\mbox{$\theta^2$,$\bar\theta^1$,$\theta^1$,$\theta^0$
terms}\Big\rangle_n = \nonumber\\
&& = \frac{1}{n}\mbox{Tr}\Big\langle - 2\theta^c\theta_c
\bar\theta^d [\bar\theta_d,*] + i\bar\theta^c (\gamma^\mu)_c{}^d
\theta_d [y_\mu^*,*]
+\mbox{$\theta^2$,$\bar\theta^1$,$\theta^1$,$\theta^0$
terms}\Big\rangle_n,
\end{eqnarray}

\noindent where $y_\nu^* = x_\nu - i\bar\theta^a
(\gamma_\nu)_a{}^b \theta_b$ is an antichiral coordinate, such
that

\begin{equation}
D_a y_\mu^* = 0.
\end{equation}

\noindent Therefore, the considered loop with an external ${\bf
V}$-line can be written as

\begin{equation}
\mbox{Tr}\Big\langle -2\theta^c\theta_c \bar\theta^d
[\bar\theta_d, \ln * ] + i \bar\theta^c (\gamma^\nu)_c{}^d
\theta_d [y_\nu^*,\ln
*]+\mbox{$\theta^2$,$\bar\theta^1$,$\theta^1$,$\theta^0$
terms}\Big\rangle.
\end{equation}

\noindent Total derivatives in this expression give 0, because
$\ln *$ does not contain $\partial_\mu/\partial^4$. Thus, the loop
gives only

\begin{equation}\label{Loop_With_V_Line}
\mbox{$\theta^2$,$\bar\theta^1$,$\theta^1$,$\theta^0$ terms}.
\end{equation}

The considered diagrams contain two loops of the matter
superfields with an attached external ${\bf V}$-line. Therefore,
in order to calculate such diagrams it is necessary to multiply
expressions (\ref{Loop_With_V_Line}), corresponding to each of
these loops. As we explained above, in order to obtain a
nontrivial result, it is necessary to have at least the second
power of $\bar\theta$ and of $\theta$. Therefore, all these terms
vanish after the multiplication and subsequent calculation of the
diagram. Thus, the sum of all such diagrams is given by an
integral of a total derivative and is equal to 0.

\subsection{External ${\bf V}$-lines are attached to
a single loop of the matter superfields} \hspace{\parindent}

Now let us consider a case in that both external $V$-lines are
attached to a single loop of the matter superfields. For
simplicity we will consider only a contribution of the field
$\phi$. (This means that the external lines are attached to the
$\phi$ loop. Other loops can certainly contain
$\widetilde\phi$-propagators.) The matter is that in the massless
limit the fields $\phi$ and $\widetilde\phi$ decouple. It is easy
to see that a contribution of the field $\widetilde\phi$ is
exactly equal to the contribution of the field $\phi$.

In the massless case we should calculate the diagrams

\vspace*{1.2cm} \hspace*{-1cm}
\begin{picture}(0,0)

\put(1.5,0.1){\includegraphics[scale=0.4]{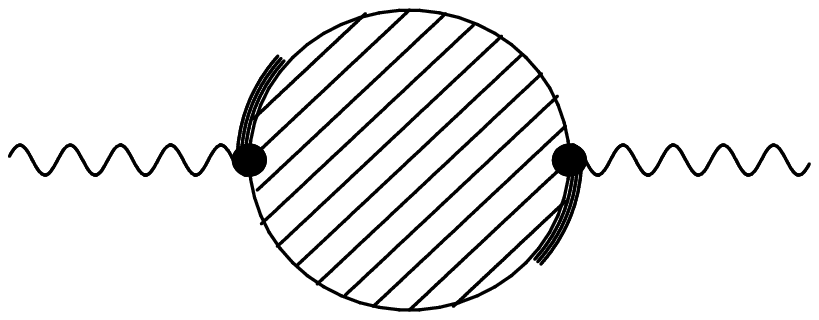}}
\put(1.4,0.2){$\theta^a \bar\theta^b$}
\put(4.2,0.2){$\theta^c\bar\theta^d$}

\put(6.5,0.1){\includegraphics[scale=0.4]{ins3.eps}}
\put(6.4,0.2){$\theta^a \bar\theta^b$}
\put(8.9,0.2){$\theta^c\theta_c\bar\theta^d$}

\put(11.5,0.1){\includegraphics[scale=0.4]{ins3.eps}}
\put(11.4,0.2){$\theta^a\theta_a \bar\theta^b$}
\put(13.9,0.2){$\theta^c\theta_c\bar\theta^d$}

\end{picture}

\vspace*{-1.7cm}
\begin{equation}
\end{equation}
\vspace*{0.1cm}

\noindent where the vertexes are given by the corresponding terms
in Eq. (\ref{Subdiagrams_Massless}).

Let us shift $\theta$-s to an arbitrary point of the loop,
commuting them with matter propagators. This gives (the
coefficients correspond to the expression (\ref{We_Calculate2});
contribution of $\widetilde\phi$ is not taken into
account)\footnote{In order to obtain the contribution of a
$\widetilde\phi$ loop it is necessary to make a substitution
$e^{2V}\to e^{-2V}$ and $*\to \widetilde *$.}

\vspace*{1.5cm}

\hspace*{4cm}
\begin{picture}(0,0)
\put(-4.5,0.1){\includegraphics[scale=0.4]{ins3.eps}}
\put(-4.6,0.2){$\theta^a \bar\theta^b$}
\put(-1.8,0.2){$\theta^c\bar\theta^d$} \put(0,0.6){${\displaystyle
= \frac{i}{64} \frac{d}{d\ln\Lambda}
\mbox{Tr}\Big\langle \theta^4 (e^{2V}-1)\frac{\bar D^2 D^2
\partial^\mu}{\partial^4} * (e^{2V}-1)\frac{\bar D^2 D^2
\partial_\mu}{\partial^4} *\Big\rangle;}$}
\end{picture}

\vspace*{-0.5cm} \hspace*{4cm}
\begin{picture}(0,0)
\put(-4.5,-1.6){\includegraphics[scale=0.4]{ins3.eps}}
\put(-4.6,-1.5){$\theta^a \bar\theta^b$}
\put(-2.1,-1.5){$\theta^c\theta_c\bar\theta^d$}

\put(0,-1.1){${\displaystyle = 2 (\gamma^\mu)_d{}^c
\frac{d}{d\ln\Lambda}
\mbox{Tr}\Big\langle\theta^4 \Big( (e^{2V}-1)\frac{\bar D^2 D_c
\partial_\mu}{16\partial^4} * (e^{2V}-1)\frac{\bar D^d D^2}{\partial^2} *
}$} \put(0,-2.2){ $ {\displaystyle + (e^{2V}-1)\frac{\bar D^2 D^2
\partial_\mu}{16\partial^4} * (e^{2V}-1)\frac{\bar D^2
D_c}{16\partial^2} * (e^{2V}-1)\frac{\bar D^d D^2}{\partial^2}
*\Big)\Big\rangle;} $}
\end{picture}

\vspace*{-0.9cm}
\hspace*{4cm}
\begin{picture}(0,0)
\put(-4.5,-4.5){\includegraphics[scale=0.4]{ins3.eps}}
\put(-4.6,-4.4){$\theta^a\theta_a \bar\theta^b$}
\put(-2.1,-4.4){$\theta^c\theta_c\bar\theta^d$}

\put(0,-4.0){ $ {\displaystyle = -2i \frac{d}{d\ln\Lambda}
\mbox{Tr}\Big\langle \theta^4
\Big( - (e^{2V}-1)\frac{\bar D_d D^2}{4\partial^2}
* (e^{2V}-1)\frac{\bar D^2}{8\partial^2}
* }$}

\put(0,-5.1){ $ {\displaystyle \times (e^{2V}-1)
\frac{\bar D^d D^2}{4\partial^2} * - (e^{2V}-1)
\frac{\bar D_d}{2\partial^2}
* (e^{2V}-1)\frac{\bar D^d D^2}{4\partial^2} * }$}

\put(-5,-6.2){ $ {\displaystyle + (e^{2V}-1)\frac{\bar D_d
D^c}{2\partial^2} * (e^{2V}-1) \frac{\bar D^2
D_c}{8\partial^2} * (e^{2V}-1)\frac{\bar D^d D^2}{4\partial^2} * +
(e^{2V}-1)\frac{\bar D_d D^2}{4\partial^2}
* (e^{2V}-1) }$}

\put(-5,-7.3){ $ {\displaystyle \times\frac{\bar D^2 D^c}{8\partial^2}
* (e^{2V}-1)\frac{\bar D^2
D_c}{8\partial^2} * (e^{2V}-1)\frac{\bar D^d
D^2}{4\partial^2} * \Big)\Big\rangle. }$}
\end{picture}

\vspace*{7.9cm}

We will start with the calculation of the following sum of
diagrams:

\vspace*{1.3cm}

\begin{picture}(0,0)
\put(3.3,0.2){\includegraphics[scale=0.4]{ins3.eps}}
\put(3.3,0.3){$\theta^a \bar\theta^b$}
\put(5.9,0.3){$\theta^c\bar\theta^d$}

\put(7.5,0.7){$+$} \put(8.2,0.7){${\displaystyle\frac{1}{2}}$}

\put(9.3,0.2){\includegraphics[scale=0.4]{ins3.eps}}
\put(9.3,0.3){$\theta^a \bar\theta^b$}
\put(11.7,0.3){$\theta^c\theta_c\bar\theta^d$}

\end{picture}

\vspace*{-1.7cm}
\begin{equation}
\end{equation}
\vspace*{0.01cm}

\noindent Using the identity

\begin{equation}\label{Delta_Commutator}
[x^\mu,\frac{\partial_\mu}{\partial^4}] =
[-i\frac{\partial}{\partial p_\mu}, -\frac{ip^\mu}{p^4}] = -2\pi^2
\delta^4(p_E) = -2\pi^2 i \delta^4(p)
\end{equation}

\noindent after simple algebraic transformations we obtain

\begin{eqnarray}\label{First_Sum_Total_Derivative}
&& 2i\frac{d}{d\ln\Lambda}\mbox{Tr}\Big\langle\theta^4 \Big(\frac{i\pi^2}{8}
* (e^{2V}-1)\bar D^2 D^2 \delta^4(\partial_\alpha) +
\Big[y_\mu^*, (e^{2V}-1)\frac{\bar D^2
D^2\partial^\mu}{16\partial^4} *\Big]\Big)\Big\rangle\quad \nonumber\\
&& = - \frac{d}{d\ln\Lambda}
\mbox{Tr}\Big\langle \frac{\pi^2}{4} \theta^4 * (e^{2V}-1)\bar D^2
D^2 \delta^4(\partial_\alpha)\Big\rangle.\qquad
\end{eqnarray}

\noindent Terms proportional to the $\delta$-function will be
calculated in the next section. (So far we have not yet found all
such terms.) We will see that they give a part of the
$\beta$-function proportional to the anomalous dimension.

Now let us calculate the diagrams

\vspace*{1.3cm}

\begin{picture}(0,0)
\put(2.5,0.2){\includegraphics[scale=0.4]{ins3.eps}}
\put(2.3,0.3){$\theta^a\theta_a \bar\theta^b$}
\put(5.0,0.3){$\theta^c\theta_c\bar\theta^d$}

\put(7,0.75){$+$} \put(7.5,0.75){${\displaystyle\frac{1}{2}}$}

\put(8.7,0.2){\includegraphics[scale=0.4]{ins3.eps}}
\put(8.7,0.3){$\theta^a \bar\theta^b$}
\put(11.2,0.3){$\theta^c\theta_c\bar\theta^d$}

\end{picture}

\vspace*{-1.7cm}
\begin{equation}
\vphantom{\int\limits_p}
\end{equation}

\noindent This sum can be written as

\begin{eqnarray}\label{Second_Sum}
&& \frac{d}{d\ln\Lambda} \mbox{Tr}\Big\langle\theta^4
\Big((\gamma^\mu)_d{}^c\Big[y_\mu^*,(I_1)_c
* (\bar I_1)^d *\Big] + 2 (\gamma^\mu)_d{}^c (I_0)
\frac{\partial_\mu}{\partial^2} * \Big((I_1)_c * (\bar I_1)^d * \qquad\nonumber\\
&& + (\bar I_1)^d * (I_1)_c *\Big) -2i\Big( 2(I_2) * (\bar I_1)^d
* (\bar I_1)_d * +2 (I_2)_d{}^c *
(I_1)_c * (\bar I_1)^d * \nonumber\\
&&+4 (\bar I_1)_d * (I_1)^c * (I_1)_c * (\bar I_1)^d *
\Big)\Big)\Big\rangle  +\mbox{terms proportional to a
$\delta$-function}.\qquad
\end{eqnarray}

\noindent (Terms proportional to a $\delta$-function will be
calculated later.)

In order to present this expression as an integral of a total
derivative we will use the identity

\begin{eqnarray}\label{Triple_Identity}
&& \mbox{Tr}\Big(\theta^4 \Big( (\gamma^\mu)^{ab} [y_\mu^*,A]
[\bar\theta_b, B\}[\theta_a, C\} + (\gamma_\mu)^{ab} (-1)^{P_A}
[\theta_a,B\} [\bar\theta_b, C\} [y_\mu^*,A] \vphantom{\frac{1}{2}}\nonumber\\
&&\qquad\qquad\qquad\qquad -4i [\theta^a,[\theta_a, A\}\}
[\bar\theta^b,B\}[\bar\theta_b,C\}\Big)\Big) +\mbox{cyclic
perm. of $A$, $B$, $C$}\qquad\vphantom{\frac{1}{2}}\nonumber\\
&& = \frac{1}{3}\mbox{Tr}\Big(\theta^4 (\gamma^\mu)^{ab}
\Big[y_\mu^*, A [\bar\theta_b,B\} [\theta_a, C\} + (-1)^{P_A}
[\theta_a,B\}[\bar\theta_b, C\}A \Big]\nonumber\\
&&\qquad\qquad\qquad\qquad\qquad\qquad\qquad\qquad\qquad\qquad\qquad
+\mbox{cyclic perm. of $A$, $B$, $C$},\qquad\vphantom{\frac{1}{2}}
\end{eqnarray}

\noindent which was proved in Ref. \cite{FactorizationHEP}. For
the completeness we also present this proof in the Appendix
\ref{Appendix_Identity}. Here $A$, $B$, and $C$ are arbitrary
differential operators, constructed from the supersymmetric
covariant derivatives, which do not explicitly depend on $\theta$,
and $P_X$ is a  Grassmannian parity of $X$.

Qualitative arguments presented in Ref. \cite{FactorizationHEP}
allow to suggest that expression (\ref{Second_Sum}) for a diagram
with $n$ vertexes on the considered matter loop can be written in
the following form:

\begin{eqnarray}\label{Second_Sum_Rewritten}
&& \frac{d}{d\ln\Lambda} \mbox{Tr}\Big\langle\theta^4
(\gamma^\mu)_d{}^c\Big[y_\mu^*,(I_1)_c
* (\bar I_1)^d *\Big]\Big\rangle_n
- \frac{6}{n(n+1)(n+2)}\frac{d}{d\ln\Lambda}\mbox{Tr}\Big\langle \theta^4 \Big(
\nonumber\\
&& + (\gamma^\mu)^{ab} [y_\mu^*,*^3][\bar\theta_b,*][\theta_a,I_0]
+ (\gamma^\mu)^{ab} [\theta_a,*^3][\bar\theta_b,*][y_\mu^*,I_0]
-4i \{\theta^a,[\theta_a,*^3]\} [\bar\theta^b, *][\bar\theta_b,
I_0]
\vphantom{\Big(}\nonumber\\
&& + (\gamma^\mu)^{ab} [y_\mu^*,*][\bar\theta_b,I_0][\theta_a,*^3]
+ (\gamma^\mu)^{ab} [\theta_a,*][\bar\theta_b,I_0][y_\mu^*,*^3]
-4i \{\theta^a,[\theta_a,*]\} [\bar\theta^b, I_0][\bar\theta_b,
*^3]
\vphantom{\Big(}\nonumber\\
&& + (\gamma^\mu)^{ab} [y_\mu^*,I_0][\bar\theta_b,*^3][\theta_a,*]
+ (\gamma^\mu)^{ab} [\theta_a,I_0][\bar\theta_b,*^3][y_\mu^*,*]
-4i \{\theta^a,[\theta_a,I_0]\} [\bar\theta^b, *^3][\bar\theta_b,
*]\qquad
\vphantom{\Big(}\nonumber\\
&& + (\gamma^\mu)^{ab} [y_\mu^*,*][\bar\theta_b,*][\theta_a,*] +
(\gamma^\mu)^{ab} [\theta_a,*][\bar\theta_b,*][y_\mu^*,*] -4i
\{\theta^a,[\theta_a,*]\} [\bar\theta^b, *][\bar\theta_b, *]
\Big)\Big\rangle_n
\end{eqnarray}

\noindent In order to prove this, it is necessary to calculate all
commutators and take into account Eq. (\ref{Star_Coefficients}).
For example, if $n=a+b+c+3$, then

\begin{eqnarray}
&& A (*)_a B (*)_b C (*)_c = \frac{6}{n(n+1)(n+2)} \Big(A (*^4)_a
B (*)_b C (*)_c
+ A (*)_a B (*^4)_b C (*)_c \nonumber\\
&& + A (*)_a B (*)_b C (*^4)_c + A (*^3)_a B (*^2)_b C (*)_c + A
(*^2)_a B (*^3)_b C (*)_c + A (*^3)_a B (*)_b C (*^2)_c
\vphantom{\frac{1}{2}}\nonumber\\
&& + A (*^2)_a B (*)_b C (*^3)_c + A (*)_a B (*^2)_b C (*^3)_c + A
(*)_a B (*^3)_b C (*^2)_c
+ A (*^2)_a B (*^2)_b C (*^2)_c \Big),\vphantom{\frac{1}{2}}\nonumber\\
\end{eqnarray}

\noindent because

\begin{eqnarray}
&& \frac{1}{6} (a+b+c+3) (a+b+c+4) (a+b+c+5)
= \frac{1}{6}(a+1)(a+2)(a+3)\nonumber\\
&& + \frac{1}{6}(b+1)(b+2)(b+3) + \frac{1}{6}(c+1)(c+2)(c+3)
+ \frac{1}{2} (a+1)(a+2)(b+1) \nonumber\\
&& + \frac{1}{2}(b+1)(b+2)(a+1) + \frac{1}{2}(a+1)(a+2)(c+1)
+ \frac{1}{2}(c+1)(c+2)(a+1)\nonumber\\
&& + \frac{1}{2}(c+1)(c+2)(b+1) + \frac{1}{2}(b+1)(b+2)(c+1) +
(a+1)(b+1)(c+1).
\end{eqnarray}

\noindent (A similar, but larger identity can be written for the
term with four $*$ in Eq. (\ref{Second_Sum}).)

Applying identity (\ref{Triple_Identity}) to
(\ref{Second_Sum_Rewritten}), we present expression
(\ref{Second_Sum}) as an integral of a total derivative:

\begin{eqnarray}\label{Second_Sum_Total_Derivative1}
&& \frac{d}{d\ln\Lambda} \mbox{Tr}\Big\langle\theta^4
(\gamma^\mu)_d{}^c\Big[y_\mu^*,(I_1)_c
* (\bar I_1)^d *\Big]\Big\rangle_n
- \frac{2}{n(n+1)(n+2)}\frac{d}{d\ln\Lambda}\mbox{Tr}\Big\langle
\theta^4 (\gamma^\mu)^{ab}\qquad
\nonumber\\
&& \times\Big[ y_\mu^*, *^3 [\bar\theta_b,*][\theta_a,I_0] +
[\theta_a,*^3][\bar\theta_b,*] I_0 +
*[\bar\theta_b,I_0][\theta_a,*^3]  +
[\theta_a,*][\bar\theta_b,I_0]*^3
\nonumber\\
&& + I_0[\bar\theta_b,*^3] [\theta_a,*] +
[\theta_a,I_0][\bar\theta_b,*^3]* + *[\bar\theta_b,*][\theta_a,*]
+ [\theta_a,*][\bar\theta_b,*]*
\Big]\Big\rangle_n\\
&& \qquad\qquad\qquad\qquad\qquad\qquad\qquad\qquad\quad
+\mbox{terms, proportional to a $\delta$-function}.\nonumber
\end{eqnarray}

\noindent Calculating the commutators with $\theta$ and
$\bar\theta$ we obtain

\begin{eqnarray}\label{Second_Sum_Total_Derivative2}
&& \frac{d}{d\ln\Lambda}\mbox{Tr}\Big\langle\theta^4
(\gamma^\mu)_d{}^c\Big[y_\mu^*,(I_1)_c
* (\bar I_1)^d *\Big]\Big\rangle_n
- \frac{2(\gamma^\mu)^{ab}}{n(n+1)(n+2)}\mbox{Tr}
\frac{d}{d\ln\Lambda} \Big\langle
\theta^4  \Big[ y_\mu^*, *^4 (\bar I_1)_b *
\nonumber\\
&& \times (I_1)_a + \Big(*^3
(I_1)_a * + *^2 (I_1)_a *^2
+ *(I_1)_a *^3\Big)* (\bar I_1)_b* I_0 + *(\bar I_1)_b \Big(*^3 (I_1)_a *
+ *^2
\nonumber\\
&& \times (I_1)_a *^2 + *(I_1)_a *^3\Big) + * (I_1)_a * (\bar I_1)_b *^3
+ I_0\Big(*^3 (\bar I_1)_b * + *^2 (\bar I_1)_b *^2 + *(\bar
I_1)_b *^3\Big)\nonumber\\
&& \times *(I_1)_a * + (I_1)_a \Big(*^3 (\bar I_1)_b * + *^2
(\bar I_1)_b *^2
+ *(\bar I_1)_b *^3\Big)* + *^2 (\bar I_1)_b *^2 (I_1)_a* +
*(I_1)_a\nonumber\\
&& \times *^2 (\bar I_1)_b*^2 \Big]\Big\rangle_n +\mbox{terms,
proportional to a $\delta$-function}.
\end{eqnarray}

Thus, the sum of all remaining diagrams is also given by an
integral of a total derivative.

In order to simplify the obtained expressions we derive an
identity, which corresponds to shifting a loop momentum in an
integral of a total derivative. For this purpose let us formally
assume that $y_\mu^*$ and $\theta^4$ do not commute. Then due to
the Jacobi identity

\begin{equation}
[[\theta^4,y_\mu^*],A] = [\theta^4,[y_\mu^*,A]] -
[y_\mu^*,[\theta^4,A]].
\end{equation}

\noindent As earlier, we assume that $A$ is a differential
operator constructed from the supersymmetric covariant
derivatives. As a consequence

\begin{equation}
[y_\mu^*,A] = -2i (\gamma_\mu)^{ab}\theta_a [\bar\theta_b,A\} +
O(\theta^0).
\end{equation}

\noindent Therefore,

\begin{equation}
[[\theta^4,y_\mu^*],A] = -2i (\gamma_\mu)^{ab} [\theta^4, \theta_a
[\bar\theta_b,A\}] + 2i (\gamma_\mu)^{ab}\theta_a
[\bar\theta_b,[\theta^4,A]\} + O(\theta^3) =O(\theta^3).
\end{equation}

\noindent (Terms that do not contain $\theta^4$ vanish after
integration over $d^4\theta$.) So, without using the relation
$[y_\mu^*,\theta^4] = 0$ we obtained

\begin{equation}
[[\theta^4,y_\mu^*],A] = O(\theta^3).
\end{equation}

\noindent Because the operation $\mbox{Tr}$ includes the
integration over $d^4\theta$, this means that it is possible to
make cyclic permutations ($P_A=P_B$)

\begin{equation}\label{Cycle_Permutation}
\mbox{Tr} \Big\langle \theta^4 [y_\mu^*,AB] \Big\rangle =
\mbox{Tr} \Big\langle [\theta^4, y_\mu^*] AB \Big\rangle =
\mbox{Tr} \Big\langle A [\theta^4, y_\mu^*] B \Big\rangle =
(-1)^{P_A}\mbox{Tr} \Big\langle \theta^4 [y_\mu^*,BA] \Big\rangle.
\end{equation}

\noindent Actually this corresponds to shifts of the loop momentum
in an integral of a total derivative. Because the integrals are
well defined, such shifts do not change the integral.

Taking into account the possibility of making such cyclic
permutations, one can simplify expression
(\ref{Second_Sum_Total_Derivative2}). In Appendix
\ref{Appendix_Coefficient} we prove that it can be written as

\begin{eqnarray}\label{Second_Sum_Total_Derivative3}
&& - \sum\limits_{a+b+2=n} \frac{2(b+1)(\gamma^\mu)^{cd}}{n}
\frac{d}{d\ln\Lambda}
\mbox{Tr}\Big\langle \theta^4 \Big[ y_\mu^*, (I_1)_c (*)_a (\bar
I_1)_d (*)_b \Big]\Big\rangle
\nonumber\\
&& +\mbox{terms proportional to a $\delta$-function}.
\end{eqnarray}

\noindent Collecting all results we obtain the following
expression for a $\phi$-contribution to (\ref{We_Calculate2}):

\begin{eqnarray}\label{Phi_Contribution}
&& \frac{d}{d\ln\Lambda}\mbox{Tr}\Big\langle \theta^4
\Big[y_\mu^*, 2i(e^{2V}-1)\frac{\bar D^2 D^2
\partial_\mu}{16\partial^4} *
- \sum\limits_{a+b+2=n} \frac{2(b+1)(\gamma^\mu)^{cd}}{n} (I_1)_c
(*)_a (\bar I_1)_d (*)_b \Big]\Big\rangle_n
\nonumber\\
&& +\mbox{terms proportional to a $\delta$-function}.
\end{eqnarray}

\noindent The terms written explicitly are certainly equal to 0.
The exact NSVZ $\beta$-function is obtained from the terms
proportional to a $\delta$-function, which are calculated in the
next section.

Taking into account a possibility of making cyclic permutations in
the expression which is commuted with $y_\mu^*$,
(\ref{Phi_Contribution}) can be rewritten as

\begin{eqnarray}
&& \frac{d}{d\ln\Lambda} \frac{1}{n} \mbox{Tr}\Big\langle \theta^4
\Big[y_\mu^*, * 2i(e^{2V}-1)\frac{\bar D^2 D^2
\partial_\mu}{16\partial^4} *
- 2 (\gamma^\mu)^{cd} * (I_1)_c * (\bar I_1)_d
* \Big]\Big\rangle_n
\nonumber\\
&& +\mbox{terms proportional to a
$\delta$-function}.\vphantom{\Big(}
\end{eqnarray}

\noindent It is easy to see that this expression can be presented
in the form

\begin{equation}
\frac{d}{d\ln\Lambda} \frac{i}{n} \mbox{Tr}\Big\langle \theta^4
\Big[(y^\mu)^*, [y_\mu^*, *]\Big]\Big\rangle_n +\mbox{terms
proportional to a $\delta$-function}.
\end{equation}

\noindent Thus, (taking into account contribution of
$\widetilde\phi$-loops) finally we obtain

\begin{equation}
i \frac{d}{d\ln\Lambda} \mbox{Tr}\Big\langle \theta^4
\Big[(y^\mu)^*, [y_\mu^*, \ln(* \widetilde *)]\Big]\Big\rangle
+\mbox{terms proportional to a $\delta$-function}.
\end{equation}

\section{Derivation of the NSVZ $\beta$-function}
\hspace{\parindent}\label{Section_NSVZ}

In the previous section it was found that all integrals giving the
$\beta$-function are integrals of total derivatives. However, they
are not equal to 0, because

\begin{equation}
\frac{1}{\pi^2}\int d^4q\,\frac{1}{q^2} \frac{d}{dq^2}
f(q^2) = \int\limits_{0}^\infty dq^2 \frac{d}{dq^2} f(q^2)
= f(\infty) - f(0) = - f(0) \ne 0.
\end{equation}

\noindent ($f(\infty)=0$ due to the higher derivative
regularization.) This is equivalent to taking into account terms
with a $\delta$-function. Really, let us rewrite this equality as
follows:

\begin{eqnarray}
&& \int\limits_{0}^\infty dq^2\, \frac{d}{dq^2} f(q^2) =
\frac{1}{2\pi^2} \int d^4q\,\frac{q^\mu}{q^4} \frac{\partial
f}{\partial q^\mu} = \frac{1}{2\pi^2} \int
d^4q\,\Bigg(\frac{\partial}{\partial q^\mu}\Big(\frac{q^\mu
f}{q^4}\Big) - f \frac{\partial}{\partial q^\mu}
\Big(\frac{q^\mu}{q^4}\Big) \Bigg)
\nonumber\\
&& = - \int d^4q\,\delta^4(q) f = - f(0).
\end{eqnarray}

\noindent Thus, we see that total derivatives with respect to
$q^2$ are equivalent to total derivatives with respect to $q^\mu$
plus terms proportional to $\delta^4(q)$. In the approach
described in the previous section $\delta$-functions appear, if
$y_\mu^*$ is commuted with $\partial_\mu/\partial^4$ as in Eq.
(\ref{Delta_Commutator}). In this section we calculate all such
terms.

Qualitatively, the $\delta$-function allows to perform an
integration over a momentum of the considered matter loop. This
corresponds to cutting the loop, which gives diagrams for the
two-point Green function of the matter superfield \cite{Smilga}.
For example,

\vspace*{1.2cm}
\begin{picture}(0,0)
\put(2.8,-0.5){\includegraphics[scale=0.24]{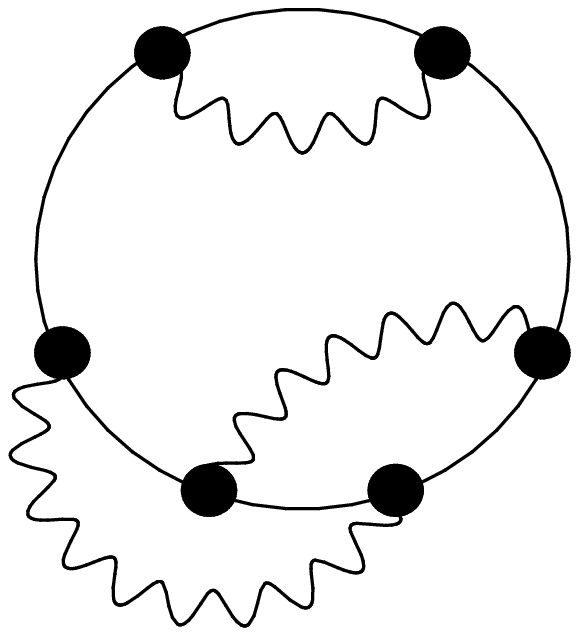}}
\put(3.5,0.3)

\put(5,0.6){\vector(1,0){1.4}}

\put(6.6,-0.5){\includegraphics[scale=0.24]{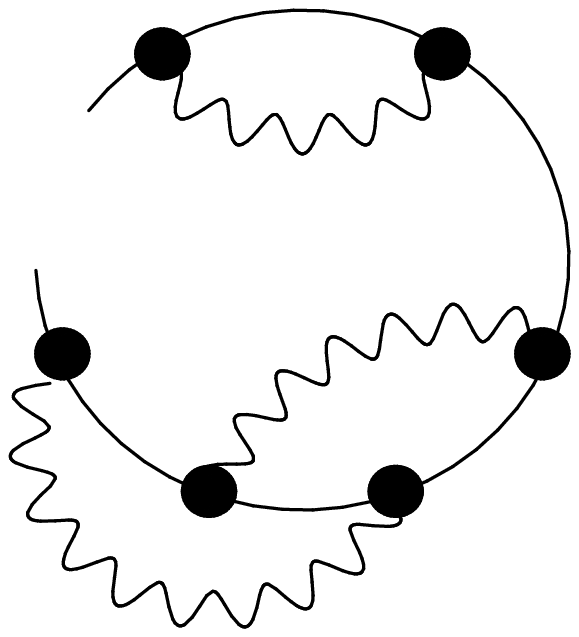}}
\put(3.5,0.3)

\put(8.8,0.6){$+$}

\put(9.1,-0.45){\includegraphics[scale=0.24]{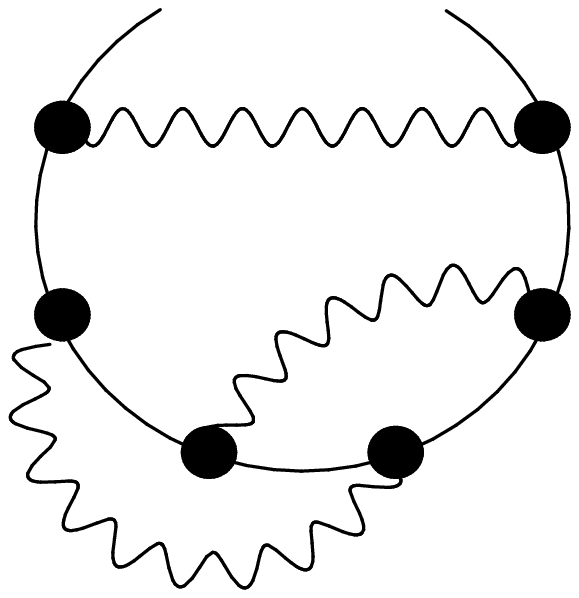}}
\put(3.5,0.3)

\put(11.2,0.6){$+\ \ \ldots$}

\end{picture}
\bigskip

Let us derive this by a rigorous method. First we consider expression
(\ref{First_Sum_Total_Derivative}). Omitting total derivatives (corresponding
to $\partial/\partial q^\mu$) we obtain

\begin{eqnarray}\label{First_Term}
&& - \frac{d}{d\ln\Lambda}\mbox{Tr}\Big\langle \frac{\pi^2}{4}
\theta^4
* (e^{2V}-1)\bar D^2 D^2 \delta^4(\partial)\Big\rangle =\nonumber\\
&& = -\frac{d}{d\ln\Lambda} \int
d^8x\,d^8y\,\delta^8_{xy}\Big\langle \frac{\pi^2}{4} \theta^4 *
(e^{2V}-1) \Big(-\frac{\bar D^2 D^2}{16\partial^2}\Big) \bar D^2
D^2 \delta^4(\partial) \delta^8_{xy} \Big\rangle =\nonumber\\
&& = - \frac{d}{d\ln\Lambda} \int
d^8x\,d^8y\,\delta^8_{xy}\Big\langle \frac{\pi^2}{4} \theta^4
(1-*) \bar D^2 D^2 \delta^4(\partial) \delta^8_{xy}
\Big\rangle =\nonumber\\
&&\qquad\qquad\qquad\qquad\qquad\quad = \frac{d}{d\ln\Lambda} \int
d^8x\,d^8y\,\delta^8_{xy}\Big\langle \frac{\pi^2}{4} \theta^4 *
\bar D^2 D^2 \delta^4(\partial) \delta^8_{xy} \Big\rangle.\qquad
\end{eqnarray}

\noindent Taking into account that

\begin{equation}
\delta^4(\partial) \delta^4(x-y) = \int \frac{d^4q}{(2\pi)^4}
\delta^4(q) e^{-iq_\alpha (x^\alpha-y^\alpha)} =
\frac{1}{(2\pi)^4}
\end{equation}

\noindent and calculating $\theta$-integrals, we obtain

\begin{equation}
- \frac{d}{d\ln\Lambda}\mbox{Tr}\Big\langle \frac{\pi^2}{4}
\theta^4
* (e^{2V}-1)\bar D^2 D^2 \delta^4(\partial)\Big\rangle
= \frac{\pi^2}{(2\pi)^4}\frac{d}{d\ln\Lambda}\int
d^4x\,d^4y\,\Big\langle * \bar D_x^2 D_x^2\delta^8_{xy}
\Big\rangle\Big|_{\theta_x= \theta_y}.
\end{equation}

\noindent Using Eq. (\ref{Formal_Anomalous_Dimension}) for the
function $G^{-1}$ this can be presented in the form

\begin{eqnarray}\label{First_Contribution}
&& \frac{\pi^2}{(2\pi)^4}\frac{d}{d\ln\Lambda}\int d^4x\,d^4y\,
G^{-1} \bar D_x^2 D_x^2\delta^8_{xy}\Big|_{\theta_x= \theta_y} =
\frac{4\pi^2}{(2\pi)^4} \frac{d}{d\ln\Lambda}\int
d^4x\,d^4y\,G^{-1}\delta^4(x-y) =\nonumber\\
&& = 4\pi^2 \frac{d}{d\ln\Lambda} G^{-1} \delta^4(p)\Big|_{p=0}.
\end{eqnarray}

\noindent This expression is not well defined. Thus, it is written
formally. However, we will see that after adding the other
contributions a well defined result is obtained.

$\delta$-functions are also present in expression
(\ref{Second_Sum_Total_Derivative3}) (or (\ref{Second_Sum})), if
the matter loop contains coinciding momentums. Really, taking into
account that

\begin{equation}
(I_1)_c = (e^{2V}-1)\frac{\bar D^2 D_c}{8\partial^2};\qquad
(\bar I_1)_d = (e^{2V}-1) \frac{\bar D_d D^2}{8\partial^2},
\end{equation}

\noindent $\partial_\mu/\partial^4$ appears due to the following
identities (It is assumed that momentums in $I_1$ and $\bar I_1$
coincide):

\begin{eqnarray}\label{I_Product1}
&& (I_1)_c \cdot (\bar I_1)_d \to \frac{\bar D^2 D_c}{8\partial^2}
\cdot \frac{\bar D_d D^2}{8\partial^2} =
\frac{i}{2}((1+\gamma_5)\gamma^\mu)_{cd}
\frac{\partial_\mu}{32\partial^4} \bar D^2 D^2;\\
\label{I_Product2} && (\bar I_1)_d \cdot (I_1)_c \to \frac{\bar
D_d D^2}{8\partial^2} \cdot \frac{\bar D^2 D_c}{8\partial^2}= -
\frac{i}{2}((1+\gamma_5)\gamma^\mu)_{cd}
\frac{\partial_\mu}{32\partial^4} D^a \bar D^2 D_a.
\end{eqnarray}

\noindent Two (or more) momentums coincide, if two cuts of the
matter loop make a diagram disconnected. An example of such a
diagram is

\begin{center}
\includegraphics[scale=0.24]{expl0.eps}
\end{center}

\noindent For analyzing such diagrams we will use the identities

\begin{eqnarray}\label{I_Identities1}
&& (\bar I_1)_b \cdot I_0 \to \frac{\bar D_b
D^2}{8\partial^2}\cdot \frac{\bar D^2 D^2}{16\partial^2} = -
\frac{\bar D_b D^2}{8\partial^2};\qquad I_0\cdot (I_1)_a \to
\frac{\bar D^2 D^2}{16\partial^2} \cdot \frac{\bar D^2
D_a}{8\partial^2} = - \frac{\bar D^2 D_a}{8\partial^2};
\nonumber\\
\label{I_Identities2} && I_0\cdot (\bar I_1)_b \to \frac{\bar D^2
D^2}{16\partial^2}\cdot \frac{\bar D_b D^2}{8\partial^2}  =
0;\qquad\qquad\ \ (I_1)_a \cdot I_0 \to \frac{\bar D^2
D_a}{8\partial^2}\cdot \frac{\bar D^2 D^2}{16\partial^2} =
0;\nonumber\\
&&\qquad\qquad\qquad\qquad I_0\cdot I_0 \to \frac{\bar D^2
D^2}{16\partial^2}\cdot \frac{\bar D^2 D^2}{16\partial^2}= -
\frac{\bar D^2 D^2}{16\partial^2}.
\end{eqnarray}

Let us assume that there are $p$ coinciding momentums $q$ in the
considered matter loop (to that the external lines are attached).
Then the corresponding diagram contributing to the (connected)
two-point Green function of the matter superfield is not 1PI and
consists of $p$ parts, connected by a single line of the matter
superfield. (If there are several groups of coinciding momentums,
each group should be considered separately.)

Let us assume that the parts of such a diagram contains $c_i$
($i=1,\ldots,p$) vertexes on the matter line (to that the external
lines are attached).\footnote{For the diagram, presented above,
$p=2$, $c_1=2$, and $c_2=4$.} We will denote expressions for these
parts by $G_1$, $G_2$, $\ldots$, $G_p$. Due to identities
(\ref{I_Identities2}) the following variants are possible:

\begin{eqnarray}\label{AB}
&& a+1 = c_1; \qquad b+1 = c_2 + c_3 + \ldots + c_p;\nonumber\\
&& a+1 = c_2; \qquad b+1 = c_1 + c_3 + \ldots + c_p;\nonumber\\
&& \qquad\qquad\qquad\qquad\ldots \qquad\qquad\qquad\nonumber\\
&& a+1 = c_p; \qquad b+1 = c_1 + c_2 +\ldots + c_{p-1},\nonumber\\
&& \quad \mbox{where}\qquad n = c_1 + c_2 +\ldots + c_p,
\end{eqnarray}

\noindent because terms with $(I_1)$-s give a nontrivial result
only if there are no lines with the momentum $q$ between $(I_1)_a$
and $(\bar I_1)_b$. (We assume that momentums in $(I_1)$ and
$(\bar I_1)$ are equal to $q$.) However, any number of such lines
can be between $(\bar I_1)_b$ and $(I_1)_a$. According to the
results of the previous section, it is necessary to calculate (and
subtract) a singular part of the expression

\begin{equation}\label{Total_Derivative}
\frac{d}{d\ln\Lambda} \mbox{Tr}\Big\langle \theta^4 \Big[y_\mu^*,
2i(e^{2V}-1)\frac{\bar D^2 D^2
\partial_\mu}{16\partial^4} *
- \sum\limits_{a+b+2=n} \frac{2(b+1)(\gamma^\mu)^{cd}}{n} (I_1)_c
(*)_a (\bar I_1)_d (*)_b \Big]\Big\rangle.
\end{equation}

\noindent A singular part of the first term has been already found
and is given by (\ref{First_Contribution}) (with the opposite
sign). For a diagram that contains a sequence of subdiagrams
$G_1$, $G_2$, $\ldots$, $G_p$ it can be written as

\begin{eqnarray}
&& (-1)^{p-1} p\cdot \frac{d}{d\ln\Lambda}
\mbox{Tr}\,\Big(\theta^4\Big[y_\mu^* ,2i \frac{\bar D^2
D^2\partial_\mu}{16\partial^4}\Big]_{\mbox{\scriptsize Singular
part}} G_1 G_2\ldots G_p\Big)\nonumber\\
&&\qquad\qquad\qquad\qquad = (-1)^{p-1} p\cdot
\frac{\pi^2}{4}\frac{d}{d\ln\Lambda} \mbox{Tr}\Big( \theta^4 G_1
G_2 \ldots G_p \bar D^2 D^2\delta^4(q) \Big).\qquad
\end{eqnarray}

\noindent The factor $p$ is present, because there are $p$
variants by which $\bar D^2 D^2\partial_\mu/16\partial^4$ can be
placed between $G_i$.

In order to calculate a singular part of the second term, we
consider

\begin{equation}
-2(\gamma^\mu)^{cd} \frac{d}{d\ln\Lambda}\sum\limits_{a+b+2=n}
\frac{(b+1)}{n} \mbox{Tr}\Big\langle \theta^4 \Big[y_\mu^*,
(I_1)_c (*)_a (\bar I_1)_d (*)_b \Big]_{\mbox{\scriptsize Singular
part}}\Big\rangle.
\end{equation}

\noindent Using a possibility of making cyclic permutations inside
the commutator and identity (\ref{I_Product2}), this expression
can be written as

\begin{eqnarray}
&& (-1)^{p-2} \frac{d}{d\ln\Lambda} \mbox{Tr}\Big\langle \theta^4
\Big[y_\mu^*, \frac{i\partial^\mu}{8\partial^4} G_1 G_2 \ldots G_p
\bar D^a D^2 D_a \Big]_{\mbox{\scriptsize Singular
part}}\Big\rangle\nonumber\\
&&\times \Bigg(\frac{c_1 + \ldots + c_{p-1}}{c_1 + c_2 + \ldots
c_n} + \frac{c_1 + \ldots + c_{p-2} + c_p}{c_1 + c_2 + \ldots c_n}
+ \ldots + \frac{c_2 + \ldots + c_p}{c_1 + c_2 + \ldots c_n}\Bigg)
\nonumber\\
&&\qquad\qquad\qquad\qquad\quad = (-1)^p (p-1)\cdot
\frac{\pi^2}{4}\frac{d}{d\ln\Lambda} \mbox{Tr}\Big(\theta^4 G_1
G_2 \ldots G_p \bar D^2 D^2\delta^4(q) \Big),\qquad
\end{eqnarray}

\noindent
because

\begin{equation}
D^a \bar D^2 D_a \delta^4(\partial) = \bar D^2 D^2 \delta^4(\partial).
\end{equation}

\noindent Therefore, a ratio of the coefficients in the first and
the second terms of Eq. (\ref{Total_Derivative}) is

\begin{equation}
- \frac{p-1}{p}.
\end{equation}

\noindent Taking a sum of both contributions we see that a
coefficient is proportional to

\begin{equation}
1 - \frac{p-1}{p} =\frac{1}{p}.
\end{equation}

\noindent The coefficient $1$ corresponds to the expansion of
$G^{-1}$ (see Eq. (\ref{First_Contribution})). Therefore, taking
into account the expansions

\begin{equation}
\ln (1-x) = -\sum\limits_{p=1}^\infty \frac{x^p}{p};\qquad
\frac{1}{1-x} = \sum\limits_{p=0}^\infty x^p,
\end{equation}

\noindent we see that singular parts of the commutators give the
contribution to Eq. (\ref{We_Calculate2})

\begin{equation}
- 4\pi^2 \delta^4(p)\Big|_{p=0} \frac{d\ln G}{d\ln\Lambda}  = -
4\pi^2 \delta^4(p)\Big|_{p=0} \frac{d}{d\ln\Lambda}\Big(\ln
(ZG)-\ln Z\Big) = -4\pi^2 \delta^4(p)\Big|_{p=0} \gamma(\alpha_0).
\end{equation}

\noindent Here the expression $\gamma(\alpha_0)$ is well defined,
unlike the corresponding expression in Eq.
(\ref{First_Contribution}). Thus, after taking into account all
contributions, the well defined result is obtained.

Diagrams with a loop of $\widetilde\phi$-fields (to that external
lines are attached) give exactly the same result. Therefore, due
to Eq. (\ref{We_Calculate2}) a $\beta$-function is given by the
sum of the one-loop contribution $\alpha^2/\pi$ and

\begin{equation}
\Delta\beta = -\frac{\alpha^2}{\pi}\gamma(\alpha).
\end{equation}

\noindent Thus, we obtain the exact NSVZ $\beta$-function

\begin{equation}
\beta(\alpha) = \frac{\alpha^2}{\pi} (1-\gamma(\alpha)).
\end{equation}

\section{Pauli--Villars contributions}
\hspace{\parindent}\label{Pauli_Villars}

Previous calculation was formal, because so far we did not take
into account contributions of the Pauli--Villars fields. However,
these contributions can be considered in a similar way.

\subsection{Summation of subdiagrams}
\hspace{\parindent}

Let us start with the summation of subdiagrams. For the
Pauli--Villars fields there are four different types of
subdiagrams. Below we will calculate elements 22, 23, 14 and 11 of
the corresponding matrix. After the substitution $V \to \theta^4$
and some algebraic transformations (omitting for simplicity
expressions for the left vertexes) they can be presented in the
following form:

\begin{picture}(0,0)
\put(-0.7,-2.0){\includegraphics[scale=0.4]{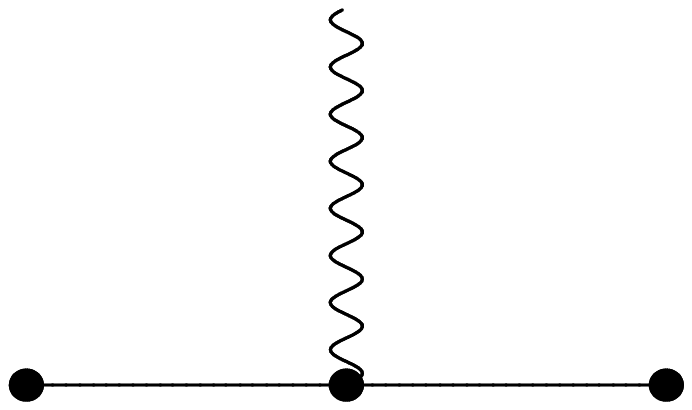}}
\put(3.3,-2.0){\includegraphics[scale=0.4]{ver01.eps}}
\put(7.3,-2.0){\includegraphics[scale=0.4]{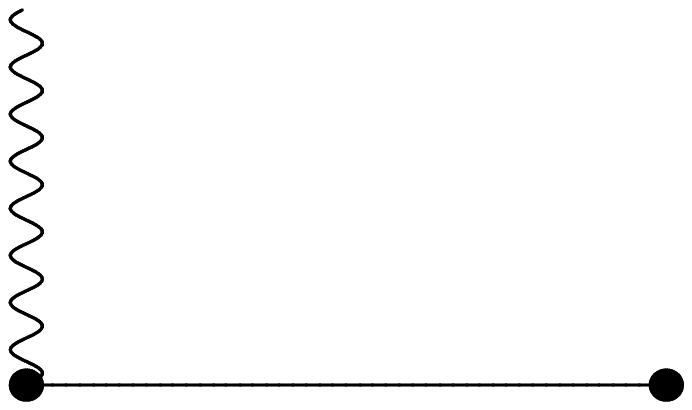}}
\put(0,-1.7){$|$} \put(1.4,-1.7){$|$} \put(4,-1.7){$|$}
\put(4.7,-1.7){$|$} \put(8.1,-1.7){$|$} \put(3,-1){$+$}
\put(6.9,-1){$+$}
\end{picture}

\vspace*{15mm}

\begin{eqnarray}\label{Subdiagram1}
&& = i \bar\theta^a (\gamma_\mu)_a{}^b \theta_b \frac{\bar D^2
D^2\partial_\mu}{8(\partial^2 + M^2)^2} - \theta^a \theta_a
\bar\theta^b \frac{\bar D_b D^2}{4(\partial^2 + M^2)} +
\mbox{terms without $\bar\theta$}\\
&& = i\bar\theta^a (\gamma_\mu)_a{}^b \theta_b \Big[y_\mu^*,
\frac{\bar D^2 D^2}{16(\partial^2 + M^2)}\Big] - 2\theta^a
\theta_a \bar\theta^b \Big[\bar\theta_b, \frac{\bar D^2
D^2}{16(\partial^2 + M^2)}\Big] +\mbox{terms without
$\bar\theta$}.\nonumber
\end{eqnarray}

\begin{picture}(0,0)
\put(-0.7,-2.0){\includegraphics[scale=0.4]{ver01.eps}}
\put(3.3,-2.0){\includegraphics[scale=0.4]{ver01.eps}}
\put(7.3,-2.0){\includegraphics[scale=0.4]{ver03.eps}}
\put(11.25,-2.05){\includegraphics[scale=0.4]{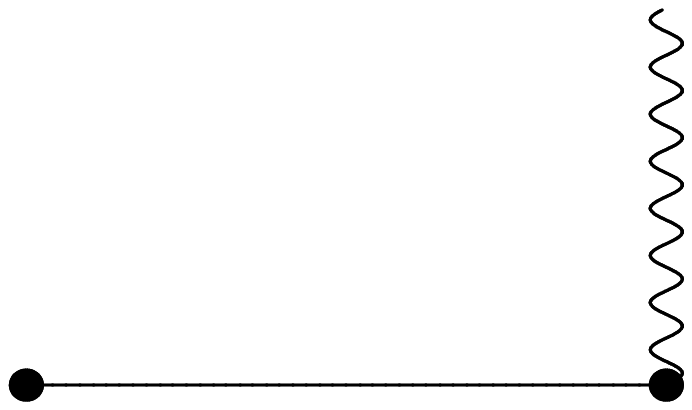}}
\put(0,-1.7){$|$} \put(1.4,-1.7){$|$} \put(4,-1.7){$|$}
\put(4.7,-1.7){$|$} \put(8.1,-1.7){$|$} \put(2.05,-1.7){$|$}
\put(6.05,-1.7){$|$} \put(10.15,-1.7){$|$} \put(3,-1){$+$}
\put(6.9,-1){$+$} \put(10.95,-1){$+$} \put(12,-1.7){$|$}
\put(14,-1.7){$|$}
\end{picture}

\vspace*{15mm}

\begin{eqnarray}\label{Subdiagram2}
&& =i\bar\theta^b (\gamma_\mu)_a{}^b \theta_a \frac{M \bar D^2
\partial_\mu}{2(\partial^2 + M^2)^2} - \theta^a \theta_a
\bar\theta^b \frac{M \bar D_b}{\partial^2 + M^2}
+\mbox{terms without $\bar\theta$} \qquad\\
&& = i\bar\theta^a (\gamma_\mu)_a{}^b \theta_b \Big[y_\mu^*,
\frac{M\bar D^2}{4(\partial^2 + M^2)}\Big] - 2\theta^a \theta_a
\bar\theta^b \Big[\bar\theta_b, \frac{M\bar D^2}{4(\partial^2 +
M^2)}\Big] +\mbox{terms without $\bar\theta$}.\quad\nonumber
\end{eqnarray}

\begin{picture}(0,0)
\put(-0.7,-2.0){\includegraphics[scale=0.4]{ver01.eps}}
\put(3.3,-2.0){\includegraphics[scale=0.4]{ver01.eps}}
\put(1.4,-1.7){$|$} \put(4.7,-1.7){$|$} \put(3,-1){$+$}
\end{picture}

\vspace*{15mm}

\begin{eqnarray}\label{Subdiagram3}
&& = -i\bar\theta^b (\gamma_\mu)_a{}^b \theta_a \frac{M D^2
\partial_\mu}{2(\partial^2 + M^2)^2}
- \bar\theta^b \frac{M \bar D_b D^2}{4(\partial^2 + M^2)^2}
+\mbox{terms without $\bar\theta$}
\\
&& = -i\bar\theta^a (\gamma_\mu)_a{}^b \theta_b \Big[y_\mu^*,
\frac{M D^2}{4(\partial^2 + M^2)}\Big] + 2\theta^a \theta_a
\bar\theta^b \Big[\bar\theta_b, \frac{M D^2}{4(\partial^2 +
M^2)}\Big] \nonumber\\
&& \qquad\qquad\qquad\qquad\qquad\qquad\qquad\qquad\qquad\quad -
\bar\theta^b \frac{M \bar D_b D^2}{4(\partial^2 + M^2)^2}
+\mbox{terms without $\bar\theta$}.\quad\nonumber
\end{eqnarray}

\begin{picture}(0,0)
\put(-0.7,-2.0){\includegraphics[scale=0.4]{ver01.eps}}
\put(3.3,-2.0){\includegraphics[scale=0.4]{ver01.eps}}
\put(7.25,-2.05){\includegraphics[scale=0.4]{ver02.eps}}
\put(0.7,-1.7){$|$} \put(5.35,-1.7){$|$} \put(2.05,-1.7){$|$}
\put(6.05,-1.7){$|$} \put(10,-1.7){$|$} \put(3,-1){$+$}
\put(6.9,-1){$+$}
\end{picture}

\vspace*{15mm}

\begin{eqnarray}\label{Subdiagram4}
&& = -i\bar\theta^b (\gamma_\mu)_a{}^b \theta_a \frac{D^2 \bar D^2
\partial_\mu}{8(\partial^2 + M^2)^2} - \theta^a \bar\theta^b
\frac{D_a \bar D_b}{\partial^2 + M^2} + \theta^a \theta_a
\bar\theta^b \frac{D^2 \bar D_b}{4(\partial^2 +
M^2)} \\
&& \qquad\qquad\qquad\qquad\qquad\qquad\ - \bar\theta^b \frac{\bar
D_b D^2 \bar D^2}{16 (\partial^2 + M^2)^2} - \bar\theta^b
\frac{\bar D_b}{\partial^2 + M^2}
+\mbox{terms without $\bar\theta$} = \nonumber\\
&& = -i\bar\theta^a (\gamma_\mu)_a{}^b \theta_b \Big[y_\mu^*,
\frac{D^2 \bar D^2}{16(\partial^2 + M^2)}\Big] + 2\theta^a
\theta_a \bar\theta^b \Big[\bar\theta_b, \frac{D^2 \bar
D^2}{16(\partial^2 + M^2)}\Big]\nonumber\\
&& \qquad\qquad\qquad\qquad\qquad\qquad\ - \bar\theta^b \frac{\bar
D_b D^2 \bar D^2}{16 (\partial^2 + M^2)^2} - \bar\theta^b
\frac{\bar D_b}{\partial^2 + M^2} +\mbox{terms without
$\bar\theta$}.\nonumber
\end{eqnarray}

\noindent The other matrix elements are calculated similarly. The
whole matrix corresponding to sums of subdiagrams
(\ref{Subdiagram1}) --- (\ref{Subdiagram4}) is written as

\begin{eqnarray}\label{Sum_Of_Massive_Subdiagrams}
&& \hspace*{-5mm}\widetilde Q\Bigg(\bar\theta^a {\cal V}\times\nonumber\\
&& \hspace*{-5mm}\left(
\begin{array}{cccc}
0 & 0 & 0 & 0 \\
{\displaystyle \frac{i(\gamma^\mu)_a{}^b D_b \bar D^2\partial_\mu
}{4(\partial^2 + M^2)^2} + \frac{\bar D_a}{\partial^2+M^2}}
& 0 & 0 & {\displaystyle\frac{M \bar D_a D^2}{4(\partial^2+M^2)^2}}\\
0 & 0 & 0 & 0 \\
0 & {\displaystyle\frac{M \bar D_a D^2}{4(\partial^2+M^2)^2}} &
{\displaystyle \frac{i(\gamma^\mu)_a{}^b D_b \bar D^2\partial_\mu
}{4(\partial^2 + M^2)^2} + \frac{\bar D_a}{\partial^2+M^2}} & 0
\end{array}
\right)\nonumber\\
&& \hspace*{-5mm} +i\bar\theta^a (\gamma^\mu)_a{}^b \theta_b
[y_\mu^*, \widetilde Q I_0] - 2\theta^a \theta_a \bar\theta^b
[\bar\theta_b,\widetilde Q I_0]+\mbox{terms without
$\bar\theta$}\Bigg),\vphantom{\Big(}
\end{eqnarray}

\noindent where

\begin{equation}
\widetilde Q \equiv \left(
\begin{array}{cccc}
-1 & 0 & 0 & 0\\
0 & 1 & 0 & 0\\
0 & 0 & 1 & 0\\
0 & 0 & 0 & -1
\end{array}
\right)
\end{equation}

\noindent satisfies the identities

\begin{equation}\label{Q_Identities}
[\widetilde Q,I_0]=0;\qquad [\widetilde Q, \star] = 0;\qquad
\widetilde Q^2 = 1.
\end{equation}

\noindent $I_0$ is defined by Eq. (\ref{Star_Definition}), and
${\cal V}$ (corresponding to the left vertexes) is given by
(\ref{Nu_Definition}). Terms without $\bar\theta$ do not
contribute to diagrams for the $\beta$-function. (The integral
over $d^4\theta$ is nontrivial only if a diagram contains
$\theta^4$ and, in particular, $\bar\theta^2$.)

\subsection{External lines are attached to different matter loops}
\hspace{\parindent}

This case is very similar to the massless one. A loop of matter
superfields to that an external line is attached is now
proportional to (for simplicity we omit $-\sum\limits_I c_I$)

\begin{eqnarray}
&& \mbox{Tr}\Big\langle i\bar\theta^c (\gamma^\mu)_c{}^d \theta_d
[y_\mu^*,\widetilde QI_0] \star -2 \theta^c\theta_c \bar\theta^d
[\bar\theta_d,\widetilde Q I_0] \star +\mbox{$\bar\theta^1$
terms}\Big\rangle_n
=\nonumber\\
&& = \frac{1}{n}\mbox{Tr}\Big\langle i\bar\theta^c
(\gamma^\mu)_c{}^d \theta_d [y_\mu^*,\widetilde Q I_0] \star^2 -
2\theta^c\theta_c \bar\theta^d [\bar\theta_d,\widetilde Q I_0]
\star^2 +\mbox{$\bar\theta^1$ terms}\Big\rangle_n.\qquad
\end{eqnarray}

\noindent After some simple (but nontrivial) algebraic transformations
this expression can be rewritten as

\begin{eqnarray}
&& \frac{1}{n}\mbox{Tr}\Big\langle \widetilde
Q\Big(-2\theta^c\theta_c \bar\theta^d \star [\bar\theta_d,I_0]
\star + i\bar\theta^c (\gamma^\nu)_c{}^d \theta_d \star
[y_\mu^*,I_0] \star\Big)
+\mbox{$\theta^2$,$\bar\theta^1$,$\theta^1$,$\theta^0$
terms}\Big\rangle_n = \quad\nonumber\\
&& = \frac{1}{n}\mbox{Tr}\Big\langle \widetilde Q\Big(-
2\theta^c\theta_c \bar\theta^d [\bar\theta_d,\star] +
i\bar\theta^c (\gamma^\mu)_c{}^d \theta_d [y_\mu^*,\star]\Big)
+\mbox{$\theta^2$,$\bar\theta^1$,$\theta^1$,$\theta^0$
terms}\Big\rangle_n.
\end{eqnarray}

\noindent Therefore, as earlier, a matter loop is given by

\begin{equation}
\mbox{Tr}\Big\langle \widetilde Q\Big(-2\theta^c\theta_c
\bar\theta^d [\bar\theta_d, \ln \star ] + i \bar\theta^c
(\gamma^\nu)_c{}^d \theta_d [y_\nu^*,\ln
\star]\Big)+\mbox{$\theta^2$,$\bar\theta^1$,$\theta^1$,$\theta^0$
terms}\Big\rangle.
\end{equation}

\noindent As in the massless case, multiplying expressions for two
such loops we obtain that all diagrams in that external lines are
attached to different matter loops are given by integrals of total
derivatives. All these integrals are evidently equal to 0.

\subsection{External lines are attached to a single matter loop}
\hspace{\parindent}

Calculation of such diagrams in the massive case has some
differences from the massless case. We construct Feynman rules in
the massive case using Eq. (\ref{Sum_Of_Massive_Subdiagrams}).
Note that they are different from the corresponding rules in the
massless case, because the expression

\begin{equation}
i\bar\theta^a (\gamma^\mu)_a{}^b \theta_b [y_\mu^*, I_0]
\end{equation}

\noindent in the massless case gives

\begin{equation}
i\bar\theta^a (\gamma^\mu)_a{}^b \theta_b \Big[y_\mu^*, \frac{\bar
D^2 D^2}{16\partial^2}\Big] = i\bar\theta^a (\gamma^\mu)_a{}^b
\theta_b \frac{\bar D^2 D^2 \partial_\mu}{8 \partial^4} -
\bar\theta^a \theta^b \theta_b \frac{\bar D_a D^2}{2\partial^2}
\end{equation}

\noindent and contains terms, proportional to $\bar\theta^a
\theta^b \theta_b$.

Also in the massive case it is necessary to take into account the
effective diagrams

\vspace*{1.5cm}
\begin{picture}(0,0)
\put(4.0,0.1){\includegraphics[scale=0.4]{ins3.eps}}
\put(3.9,0.2){$\bar\theta^b$} \put(6.7,0.2){$\theta^c \theta_c
\bar\theta^a$}
\put(8.5,-0.4){\includegraphics[scale=0.4]{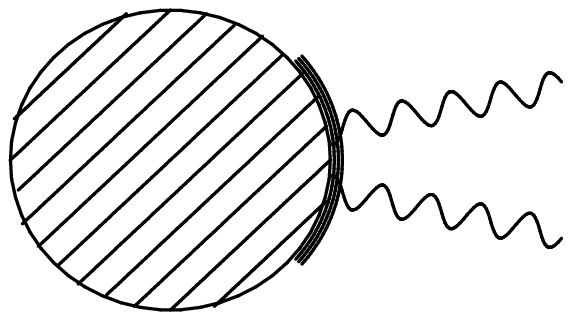}}
\end{picture}

\vspace*{-1.7cm}
\begin{equation}
\end{equation}
\vspace*{0.1cm}

\noindent The first diagram corresponds to terms, proportional to
$\bar\theta^a$ in Eq. (\ref{Sum_Of_Massive_Subdiagrams}). The
second one contains a sum of subdiagrams with two adjacent
external lines. These subdiagrams are presented in Appendix
\ref{Appendix_Subdiagrams}.

Let us now write down the results for all diagrams (again,
omitting $-\sum\limits_I c_I$ for simplicity):

The diagrams contributing in the massless case are calculated
similarly. Taking into account identities (\ref{Q_Identities}),
the result can be written as

\vspace*{1.6cm}

\hspace*{4cm}
\begin{picture}(0,0)
\put(-4.5,0.1){\includegraphics[scale=0.4]{ins3.eps}}
\put(-4.6,0.2){$\theta^a \bar\theta^b$}
\put(-1.8,0.2){$\theta^c\bar\theta^d$}
\put(-0.3,0.6){${\displaystyle = Ex_1 + \frac{3}{2}Ex_2 + 4Ex_3}$}

\put(-4.5,-1.6){\includegraphics[scale=0.4]{ins3.eps}}
\put(-4.6,-1.5){$\theta^a \bar\theta^b$}
\put(-2.1,-1.5){$\theta^c\theta_c\bar\theta^d$}
\put(-0.3,-1.1){$= -Ex_2 - 4Ex_3$}

\put(-4.5,-3.3){\includegraphics[scale=0.4]{ins3.eps}}
\put(-4.6,-3.2){$\theta^a\theta_a \bar\theta^b$}
\put(-2.1,-3.2){$\theta^c\theta_c\bar\theta^d$}

\put(-0.3,-2.8){ $=Ex_3$}

\put(4,-1.2){$\left.\vphantom{\begin{array}{c}a\\a\\a\\a\\a\\a\\a\\a\\
\end{array}}\right\}
{\displaystyle = Ex_1+ \frac{1}{2}Ex_2+Ex_3}$}

\end{picture}

\vspace*{4.0cm}

\noindent where

\begin{eqnarray}
&&\hspace*{-5mm} Ex_1 = \frac{i}{2} \frac{d}{d\ln\Lambda}
\mbox{Tr}\Big\langle \theta^4 [y_\mu^*,I_0]
\star [y_\mu^*,I_0] \star\Big\rangle;\nonumber\\
&&\hspace*{-5mm} Ex_2 = 2(\gamma^\mu)_d{}^c \frac{d}{d\ln\Lambda}
\mbox{Tr}\Big\langle\theta^4 \Big( [y_\mu^*, (I_1)_c] \star (\bar
I_1)^d \star + [y_\mu^*,I_0]
\star (I_1)_c \star (\bar I_1)^d \star\Big)\Big\rangle;\nonumber\\
&&\hspace*{-5mm} Ex_3 = -2i \frac{d}{d\ln\Lambda}
\mbox{Tr}\Big\langle \theta^4 \Big( - (\bar I_1)_d \star (I_2)
\star (\bar I_1)^d \star - (\bar I_3)_d \star (\bar I_1)^d \star +
2 (I_2)_d{}^c \star (I_1)_c \star
(\bar I_1)^d \star\nonumber\\
&& + 2 (\bar I_1)_d \star (I_1)^c \star (I_1)_c \star (\bar I_1)^d
\star \Big)\Big\rangle.
\end{eqnarray}

\noindent
Taking into account that

\begin{equation}
\star[\bar\theta_a,I_0]\star = [\bar\theta^a,\star]; \qquad
\mbox{Tr}(A[\bar\theta^a, B]) = \mbox{Tr}(\{A,\bar\theta^a\} B)
\end{equation}

\noindent and using Eq. (\ref{Q_Identities}), we also obtain

\begin{eqnarray}\label{Additional_Diagram}
&&
\begin{picture}(0,0)
\put(-0.3,-0.5){\includegraphics[scale=0.4]{ins3.eps}}
\put(-0.4,-0.4){$\bar\theta^b$} \put(2.4,-0.4){$\theta^c \theta_c
\bar\theta^a$}
\end{picture}
\hspace*{3.6cm} = -i\frac{d}{d\ln\Lambda}
\mbox{Tr}\Big\langle\theta^4 \star {\cal V} \vphantom{\int\limits_p} \\
&& \hspace*{-10mm}\left(
\begin{array}{cccc}
0 & 0 & 0 & 0\\
{\displaystyle \frac{4}{\partial^2 + M^2}
-\frac{i(\gamma^\mu)^{ab} D_a \bar D_b
\partial_\mu}{(\partial^2 + M^2)^2}}
& 0 & 0 & {\displaystyle \frac{M D^2}{(\partial^2 + M^2)^2}}\\
0 & 0 & 0 & 0\\
0 & {\displaystyle \frac{M D^2}{(\partial^2 + M^2)^2}} &
{\displaystyle \frac{4}{\partial^2 + M^2}
-\frac{i(\gamma^\mu)^{ab} D_a \bar D_b
\partial_\mu}{(\partial^2 + M^2)^2}} & 0
\end{array}
\right)\Big\rangle.\nonumber
\end{eqnarray}

The last diagram is calculated using expressions for the
subdiagrams with two adjacent external lines, presented in
Appendix \ref{Appendix_Subdiagrams}. The result is

\begin{eqnarray}\label{Two_Vertex_Diagrams}
&&\qquad \begin{picture}(0,0)
\put(-3,-1.0){\includegraphics[scale=0.4]{ins5.eps}}
\end{picture}
= -i \frac{d}{d\ln\Lambda}
\mbox{Tr}\Big\langle \theta^4 \star \frac{4M^2}{(\partial^2 +
M^2)^2} I_0\Big\rangle\vphantom{\int\limits_p}\nonumber\\
&&\qquad\qquad + i\frac{d}{d\ln\Lambda}
\mbox{Tr}\Big\langle \theta^4 \star {\cal V}\left(
\begin{array}{cccc}
0 & 0 & 0 & 0\\
0 & 0 & 0 &
{\displaystyle \frac{M D^2}{(\partial^2+M^2)^2}}\\
0 & 0 & 0 & 0\\
0 & {\displaystyle \frac{M D^2}{(\partial^2+M^2)^2}} & 0 & 0
\end{array}
\right)\Big\rangle.\qquad
\end{eqnarray}

Now let us find a sum of these diagrams. Similar to
the massless case

\begin{eqnarray}\label{Last_Diagram}
&&\hspace*{-6mm} Ex_1 = \frac{i}{2} \frac{d}{d\ln\Lambda}
\mbox{Tr}\Big\langle\theta^4 [y_\mu^*, [y_\mu^*,I_0] \star] -
\theta^4 [y_\mu^*, [y_\mu^*,I_0]] \star \Big\rangle = i
\frac{d}{d\ln\Lambda} \mbox{Tr}\Big\langle \theta^4 \frac{4
M^2}{(\partial^2 + M^2)^2}
I_0 \star\Big\rangle  \nonumber\\
&&\hspace*{-6mm} - i\frac{d}{d\ln\Lambda} \mbox{Tr}\Big\langle
\theta^4 {\cal V}\left(
\begin{array}{cccc}
0 & 0 & 0 & 0\\
{\displaystyle \frac{i (\gamma^\mu)^{ab} D_a \bar D_b
\partial_\mu}{(\partial^2 + M^2)^2} -
\frac{4}{\partial^2 + M^2}} & 0 & 0 & 0 \\
0 & 0 & 0 & 0\\
0 & 0 & {\displaystyle \frac{i (\gamma^\mu)^{ab} D_a \bar D_b
\partial_\mu}{(\partial^2 + M^2)^2} - \frac{4}{\partial^2 + M^2}} & 0
\end{array}
\right) *\Big\rangle,\nonumber\\
\end{eqnarray}

\noindent where we take into account that

\begin{eqnarray}
&& \theta^4 \Big[y_\mu^*,\Big[y_\mu^*,\frac{D^2 \bar
D^2}{16(\partial^2 + M^2)}\Big]\Big] = \theta^4\Bigg(-\frac{M^2
D^2\bar D^2}{2(\partial^2 + M^2)^3} + \frac{i D^2
(\gamma^\mu)^{ab}\theta_a \bar D_b\partial_\mu}{(\partial^2 +
M^2)^2} + \frac{2 D^2 \theta^a\theta_a}{\partial^2 +
M^2}\Bigg)\quad
\nonumber\\
&& = \theta^4\Bigg(-\frac{M^2 D^2\bar D^2}{2(\partial^2 + M^2)^3}
+ \frac{2 i (\gamma^\mu)^{ab} D_a \bar
D_b\partial_\mu}{(\partial^2 + M^2)^2} - \frac{8}{\partial^2 +
M^2}\Bigg).
\end{eqnarray}

\noindent Singularities, giving $\delta$-functions, are certainly
absent in the massive case.

Summing (\ref{Additional_Diagram}), (\ref{Two_Vertex_Diagrams}),
and (\ref{Last_Diagram}) we obtain

\begin{equation}\label{First_Massive_Total_Derivative}
\frac{i}{2} \frac{d}{d\ln\Lambda}
\mbox{Tr}\Big\langle\theta^4 [y_\mu^*, [y_\mu^*,I_0]
\star]\Big\rangle  = 0.
\end{equation}

Moreover, exactly as in the massless case for diagrams with $n$
vertexes on the matter loop

\begin{eqnarray}\label{Second_Massive_Total_Derivative}
&&\hspace*{-2mm} \frac{1}{2}Ex_2 + Ex_3 = \frac{d}{d\ln\Lambda}
\mbox{Tr}\Big\langle
\theta^4 \Big[y_\mu^*, -\frac{1}{2}(\gamma^\mu)^{cd} (I_1)_c \star
(\bar I_1)_d \star\Big]\Big\rangle_n -
\frac{1}{n(n+1)(n+2)} \frac{d}{d\ln\Lambda} \nonumber\\
&&\hspace*{-2mm} \times  \mbox{Tr}\Big\langle \theta^4
(\gamma^\mu)^{cd} \Big[ y_\mu^*, \star^3
[\bar\theta_d,\star][\theta_c,I_0] +
[\theta_c,\star^3][\bar\theta_d,\star] I_0 +
\star[\bar\theta_d,I_0][\theta_c,\star^3] +
[\theta_c,\star][\bar\theta_d,I_0]\star^3 \vphantom{\frac{1}{2}}
\qquad\nonumber\\
&&\hspace*{-2mm} +
I_0[\bar\theta_d,\star^3][\theta_c,\star]
+ [\theta_c,I_0][\bar\theta_d,\star^3]\star +
\star[\bar\theta_d,\star][\theta_c,\star] +
[\theta_c,\star][\bar\theta_d,\star]\star \Big]\Big\rangle_n =
0.\vphantom{\frac{1}{2}}
\end{eqnarray}

\noindent (It is evident that there are no terms proportional to a
$\delta$-function in this case.) Taking into account a possibility
of making cyclic permutations (\ref{Cycle_Permutation}), we obtain
that the contribution of the Pauli--Villars fields is given by the
following integral of a total derivative (for diagrams with $n$
vertexes on the matter loop to that external lines are attached):

\begin{equation}\label{Total_Derivative_Pauli_Villars}
\frac{d}{d\ln\Lambda} \mbox{Tr}\Big\langle\theta^4
\Big[y_\mu^*, \frac{i}{2} [y_\mu^*,I_0] \star
- \sum\limits_{a+b+2=n} \frac{(b+1)(\gamma^\mu)^{cd}}{n} (I_1)_c
(\star)_a (\bar I_1)_d (\star)_b \Big]\Big\rangle_n = 0.
\end{equation}

\noindent As earlier, the left hand side can be rewritten as

\begin{equation}\label{Total_Derivative_Pauli_Villars_Ln}
\frac{i}{2}\frac{d}{d\ln\Lambda} \mbox{Tr}\Big\langle\theta^4
\Big[y_\mu^*, \Big[(y^\mu)^*,\ln(\star)\Big] \Big]\Big\rangle = 0.
\end{equation}

The final result for the sum of diagrams in that the external
${\bf V}$-lines are attached to a single loop of matter
superfields is

\begin{equation}\label{Total_Derivative_Final_Result}
i\frac{d}{d\ln\Lambda} \mbox{Tr}\Big\langle\theta^4 \Big[y_\mu^*,
\Big[(y^\mu)^*,\ln(*) +\ln(\widetilde *) -\frac{1}{2}\sum\limits_I
c_I \ln(\star_I)\Big] \Big]\Big\rangle - \mbox{terms with a
$\delta$-function},
\end{equation}

\noindent where $\star_I$ means that it is necessary to use the
mass $M_I$ in the definition of $\star$. The factorization of
integrands into double total derivatives, which follows from this
equation, agrees with the arguments presented in Ref.
\cite{Smilga}.

From Eq. (\ref{Total_Derivative_Final_Result}) it is possible to
explicitly construct an integral of a total derivative, which is
equal to the sum of Feynman diagrams. In the three-loop
approximation this is made in the following section.

\section{Three-loop verification}
\hspace{\parindent}\label{Section_Three_Loop}

In order to verify the expressions, obtained in the previous
sections, we compare them with the explicit three-loop
calculation, made in Ref. \cite{3LoopHEP} by a different method.
The result can be written as \cite{FactorizationHEP}

\begin{eqnarray}
\frac{\beta(\alpha_0)}{\alpha_0^2} = \frac{d}{d\ln\Lambda}
\Big(d^{-1}(\alpha_0,\Lambda/p) - \alpha_0^{-1}\Big)\Big|_{p=0} =
16\pi (A_1 + A_2 + A_3),
\end{eqnarray}

\noindent where\footnote{This result was also presented in Ref.
\cite{Review}, but some sign in $A_3$ were written incorrectly.}

\begin{eqnarray}\label{Three_Loop_Beta1}
&& A_1 = -\frac{1}{2}\sum\limits_I c_I \int \frac{d^4q}{(2\pi)^4}
\frac{1}{q^2} \frac{d}{dq^2}
\frac{d}{d\ln\Lambda}\Big(\ln(q^2+M_I^2) +
\frac{M_I^2}{q^2+M_I^2}\Big);\\
\vphantom{0}\nonumber\\
\label{Three_Loop_Beta2} && A_2 = -2e^2 \int \frac{d^4q}{(2\pi)^4}
\frac{1}{q^2} \frac{d}{dq^2} \frac{d}{d\ln\Lambda} \int
\frac{d^4k}{(2\pi)^4} \frac{1}{k^2 R_k^2}\Bigg(\frac{1}{(k+q)^2} -
\sum\limits_{J} c_J \frac{q^4}{(q^2+M_J^2)^2}
\nonumber\\
&& \times\frac{1}{((k+q)^2+M_J^2)} \Bigg) \Bigg[ R_k
\Big(1+\frac{e^2}{4\pi^2} \ln\frac{\Lambda}{\mu}\Big) - \int
\frac{d^4t}{(2\pi)^4}\,\frac{2 e^2}{t^2 (k+t)^2}
\nonumber\\
&&  + \sum\limits_I c_I \int \frac{d^4t}{(2\pi)^4}\,
\frac{2 e^2}{(t^2+M_I^2) ((k+t)^2+M_I^2)} \Bigg];\\
\vphantom{0}\nonumber\\
\label{Three_Loop_Beta3} && A_3 = \int
\frac{d^4q}{(2\pi)^4}\,\frac{1}{q^2}\frac{d}{dq^2}
\frac{d}{d\ln\Lambda}\int \frac{d^4k}{(2\pi)^4}
\frac{d^4l}{(2\pi)^4} \frac{4 e^4}{k^2 R_k\, l^2 R_l}
\Bigg\{\frac{1}{(q+k)^2}\Bigg[- \frac{1}{2 (q+l)^2} \nonumber\\
&& + \frac{2 k^2}{(q+k+l)^2 (q+l)^2} - \frac{1}{(q+k+l)^2} \Bigg]
-\sum\limits_{I} c_I
\frac{q^4}{(q^2+M_I^2)^2} \frac{1}{((q+k)^2+M_I^2)}\times\nonumber\\
&&\times \Bigg[- \frac{1}{2 ((q+l)^2+M_I^2) } + \frac{2
k^2}{((q+k+l)^2+M_I^2) ((q+l)^2+M_I^2)}
\nonumber\\
&& - \frac{1}{((q+k+l)^2+M_I^2) } + \frac{2
M_I^2}{((q+k)^2+M_I^2)((q+k+l)^2+M_I^2)
} \nonumber\\
&&  +\frac{2 M_I^2}{(q^2+M_I^2) ((q+l)^2+M_I^2) } +
\frac{2M_I^2}{((q+l)^2+M_I^2) ((q+k+l)^2+M_I^2)}\Bigg] \Bigg\},
\end{eqnarray}

\noindent where $R_k\equiv R(k^2/\Lambda^2)$. Here $A_1$ is a
one-loop result. $A_2$ is a sum of two-loop diagrams, three-loop
diagrams with two loops of the matter superfields, and diagrams
with insertions of counterterms arising from renormalization of
the coupling constant. $A_3$ is a sum of three-loop diagrams with
a single loop of the matter superfields.

The anomalous dimension can be written as

\begin{equation}
\gamma(\alpha_0) = - \frac{d\ln Z}{d\ln\Lambda} =
\frac{d}{d\ln\Lambda}\Big(\ln(ZG)-\ln Z\Big)\Bigg|_{q=0} =
\frac{d\ln G}{d\ln\Lambda}\Bigg|_{q=0},
\end{equation}

\noindent where $\Lambda$ and the renormalized coupling constant
$\alpha$ are considered as independent variables. The two-point
Green function of the matter superfield in the two-loop
approximation is given by the following integrals
\cite{3LoopHEP,Review}:

\begin{eqnarray}\label{Two_loop_LnG}
&& \ln G = -\int \frac{d^4k}{(2\pi)^4} \frac{2 e_0^2}{k^2 R_k
(k+q)^2}\Bigg[ 1 - \frac{1}{R_k}\int
\frac{d^4t}{(2\pi)^4}\,\frac{2 e_0^2}{t^2 (k+t)^2} + \sum\limits_I
c_I \frac{1}{R_k}\int \frac{d^4t}{(2\pi)^4}\,
\nonumber\\
&& \times \frac{2 e_0^2}{(t^2+M_I^2) ((k+t)^2+M_I^2)} \Bigg] +
\int \frac{d^4k}{(2\pi)^4} \frac{d^4l}{(2\pi)^4} \frac{e_0^4}{k^2
R_k
l^2 R_l} \Bigg( -\frac{2}{(q+k)^2 (q+l)^2} \nonumber\\
&& -\frac{4}{(q+k)^2(q+k+l)^2} + \frac{8k^2-4q^2}{(q+k)^2
(q+k+l)^2 (q+l)^2} \Bigg).
\end{eqnarray}

\noindent Therefore, taking into account one-loop renormalization
of the coupling constant, we obtain

\begin{eqnarray}\label{Two_Loop_Gamma}
&&\hspace*{-5mm} \gamma(\alpha_0) = - 2e^2 \int
\frac{d^4k}{(2\pi)^4} \frac{d}{d\ln\Lambda} \frac{1}{k^4 R_k^2}
\Bigg[R_k \Big(1+\frac{e^2}{4\pi^2}\ln\frac{\Lambda}{\mu} \Big) -
\int
\frac{d^4t}{(2\pi)^4}\,\frac{2 e^2}{t^2 (k+t)^2} +\nonumber\\
&&\hspace*{-5mm} + \sum\limits_I c_I \int \frac{d^4t}{(2\pi)^4}\,
\frac{2 e^2}{(t^2+M_I^2) ((k+t)^2+M_I^2)} \Bigg] - \int
\frac{d^4k}{(2\pi)^4} \frac{d^4l}{(2\pi)^4}\,\frac{d}{d\ln
\Lambda}\frac{4 e^4 k_\mu l_\mu}{k^4 R_k\,l^4 R_l
(k+l)^2}.\nonumber\\
\end{eqnarray}

\noindent This expression is finite both in the UV and IR regions.
The UV finiteness is ensured by the regularization. The integral
is IR finite due to the differentiation with respect to
$\ln\Lambda$, which should be performed before the integration.

Taking the integrals of total derivatives in Eqs.
(\ref{Three_Loop_Beta1}) --- (\ref{Three_Loop_Beta3}) using the
identity

\begin{equation}\label{Integral_Of_Total_Derivative}
\int \frac{d^4q}{(2\pi)^4}\frac{1}{q^2} \frac{d}{dq^2} f(q^2) =
\frac{1}{16\pi^2} \Big(f(q^2=\infty) - f(q^2=0)\Big),
\end{equation}

\noindent we obtain

\begin{equation}
\frac{\beta(\alpha_0)}{\alpha_0^2} = \frac{1}{\pi}\Big(1
-\frac{d\ln G}{d\ln\Lambda}\Bigg|_{q=0}\Big) + O(\alpha_0^3).
\end{equation}

\noindent As a consequence,

\begin{equation}
\beta(\alpha) = \frac{\alpha^2}{\pi} \Big(1-\gamma(\alpha)\Big) +
O(\alpha^5).
\end{equation}

In order to verify Eq. (\ref{Two_Loop_Gamma}), it was compared
with the result of the calculation made with the dimensional
reduction. (Such calculation can be made using Eq.
(\ref{Two_loop_LnG}).) Using the standard technique of the
dimensional reduction one obtain

\begin{equation}
\gamma_{\mbox{\scriptsize DRED}}(\alpha) = -\frac{\alpha}{\pi} +
\frac{\alpha^2}{\pi^2} + O(\alpha^3).
\end{equation}

\noindent This result up to notations\footnote{In order to obtain
the results of Ref. \cite{Jones} it is necessary to set $\alpha =
g^2/4\pi$, $\gamma(\alpha) = 2\gamma(g)$, $\beta(\alpha) = g
\beta(g)/2\pi$.} agrees with the calculation made in \cite{Jones}.

From the other side, it is possible to calculate the two-point
Green function of the gauge superfield using Eqs.
(\ref{Total_Derivative}) and
(\ref{Total_Derivative_Pauli_Villars}). Taking into account that a
contribution of diagrams with a $\widetilde\phi$-loop is equal to
a contribution of diagrams with a $\phi$-loop, we obtain

\begin{eqnarray}
&& 4i
\frac{d}{d\ln\Lambda}\mbox{Tr}\Big\langle\theta^4\Big[y_\mu^*,
(e^{2V}-1)\frac{\bar D^2 D^2\partial^\mu}{16\partial^4} *\Big]
\Big\rangle +\mbox{terms with a $\delta$-function}
\nonumber\\
&& = -32\pi \int\frac{d^4q}{(2\pi)^4} \frac{1}{q^2}
\frac{d}{dq^2}\int \frac{d^4k}{(2\pi)^4}
\frac{d}{d\ln\Lambda}\frac{e^2}{k^2 R_k^2 (k+q)^2} \Bigg[ R_k
\Big(1+\frac{e^2}{4\pi^2}\ln\frac{\Lambda}{\mu}\Big)
\nonumber\\
&& - 2 e^2 \Bigg(\int \frac{d^4t}{(2\pi)^4}\,\frac{1}{t^2 (k+t)^2}
- \sum\limits_I c_I \int \frac{d^4t}{(2\pi)^4}\,
\frac{1}{(t^2+M_I^2) ((k+t)^2+M_I^2)} \Bigg)\Bigg]
\nonumber\\
&& + 64\pi \int \frac{d^4q}{(2\pi)^4} \frac{1}{q^2}
\frac{d}{dq^2}\int \frac{d^4k}{(2\pi)^4}
\frac{d^4l}{(2\pi)^4}\,\frac{d}{d\ln\Lambda} \frac{e^4}{k^2 R_k
l^2 R_l}  \Bigg( \frac{1}{(q+k)^2
(q+k+l)^2} \\
&&\qquad\qquad\qquad\qquad\qquad\qquad\qquad\qquad\qquad\qquad\quad
- \frac{(2q+k+l)^2}{(q+k)^2 (q+l)^2
(q+k+l)^2}\Bigg).\quad\nonumber
\end{eqnarray}

\noindent The last term in Eq. (\ref{Total_Derivative}) is

\begin{eqnarray}
&& -4 \sum\limits_{a+b+2=n} \frac{d}{d\ln\Lambda} \frac{b+1}{n}
(\gamma^\mu)^{cd} \mbox{Tr} \Big\langle\theta^4
\Big[y_\mu^*,(I_1)_c (*)_a (\bar I_1)_d (*)_b\Big]\Big\rangle
\nonumber\\
&& \qquad\qquad\qquad\qquad\qquad\qquad\qquad\qquad\qquad
+\mbox{terms with a $\delta$-function}\qquad\nonumber\\
&& = 32\pi \int \frac{d^4q}{(2\pi)^4} \frac{1}{q^2}
\frac{d}{dq^2}\int \frac{d^4k}{(2\pi)^4}
\frac{d^4l}{(2\pi)^4}\,\frac{d}{d\ln\Lambda} \frac{e^4}{k^2 R_k
l^2 R_l} \frac{1}{(q+k)^2 (q+l)^2}
\nonumber\\
&& + 16\pi \int \frac{d^4q}{(2\pi)^4} \frac{d^4k}{(2\pi)^4}
\frac{d^4l}{(2\pi)^4}\,\frac{d}{d\ln\Lambda} \frac{e^4}{k^2 R_k
l^2 R_l} \frac{\partial}{\partial q^\mu} \frac{(2q+k+l)^\mu}{q^2
(q+k)^2 (q+l)^2 (q+k+l)^2}.
\end{eqnarray}

\noindent Contributions of diagrams with a Pauli--Villars loop can
be calculated similarly:

\begin{eqnarray}
&& - \frac{i}{2}\sum\limits_J c_J
\frac{d}{d\ln\Lambda}\mbox{Tr}\Big\langle\theta^4\Big[y_\mu^*,
[y_\mu^*, I_0]\star\Big]
\Big\rangle =\nonumber\\
&& = 32\pi\sum\limits_J c_J \int\frac{d^4q}{(2\pi)^4}
\frac{1}{q^2} \frac{d}{dq^2} \int \frac{d^4k}{(2\pi)^4}
\frac{d}{d\ln\Lambda}\frac{e^2}{k^2 R_k^2}
\frac{q^4}{(q^2+M_J^2)^2 ((k+q)^2+M_J^2)} \Bigg[ R_k \Big(1
\nonumber\\
&& +\frac{e^2}{4\pi^2} \ln\frac{\Lambda}{\mu}\Big) - \int
\frac{d^4t}{(2\pi)^4}\,\frac{2 e^2}{t^2 (k+t)^2} + \sum\limits_I
c_I \int \frac{d^4t}{(2\pi)^4}\, \frac{2e^2}{(t^2+M_I^2)
((k+t)^2+M_I^2)} \Bigg)\Bigg]
\nonumber\\
&& - 64\pi \sum\limits_I c_I \int \frac{d^4q}{(2\pi)^4}
\frac{1}{q^2} \frac{d}{dq^2}\int \frac{d^4k}{(2\pi)^4}
\frac{d^4l}{(2\pi)^4}\,\frac{d}{d\ln\Lambda} \frac{e^4}{k^2 R_k
l^2 R_l} \frac{q^4}{(q^2+M_I^2)^2} \nonumber\\
&& \times\Bigg( \frac{1}{((q+k+l)^2 + M_I^2)((q+k)^2 + M_I^2)} -
\frac{(2q+k+l)^2 + 2 M_I^2}{((q+k)^2+M_I^2) ((q+l)^2 +M_I^2)}
\nonumber\\
&& \times\frac{1}{((q+k+l)^2 + M_I^2)} +
\frac{2M_I^2}{(q^2+M_I^2)((q+k)^2+M_I^2)((q+l)^2+M_I^2)}
\nonumber\\
&& + \frac{2M_I^2}{((q+k)^2+M_I^2)^2((q+k+l)^2+M_I^2)} \Bigg)
\nonumber\\
&& - 32\pi\sum\limits_I c_I \int \frac{d^4q}{(2\pi)^4}
\frac{1}{q^2} \frac{d}{dq^2}\int \frac{d^4k}{(2\pi)^4}
\frac{d^4l}{(2\pi)^4}\,\frac{d}{d\ln\Lambda} \frac{e^4}{k^2 R_k
l^2 R_l} \frac{q^4}{(q^2+M^2)^2 ((q+k)^2 +M_I^2)}
\nonumber\\
&& \times\frac{1}{((q+l)^2+M_I^2)} - 16\pi\sum\limits_I c_I \int
\frac{d^4q}{(2\pi)^4} \frac{d^4k}{(2\pi)^4}
\frac{d^4l}{(2\pi)^4}\,\frac{d}{d\ln\Lambda} \frac{e^4}{k^2 R_k
l^2 R_l} \frac{\partial}{\partial q^\mu}
\frac{1}{(q^2 + M_I^2)} \nonumber\\
&& \times \frac{(2q+k+l)^\mu}{((q+k)^2 + M_I^2)
((q+l)^2+M_I^2)((q+k+l)^2+M_I^2)};\\
\vphantom{\Big(}\nonumber\\
&& \sum\limits_I c_I \sum\limits_{a+b+2=n} \frac{d}{d\ln\Lambda}
\frac{b+1}{n} (\gamma^\mu)^{cd} \mbox{Tr} \Big\langle\theta^4
\Big[y_\mu^*,(I_1)_c (\star)_a (\bar I_1)_d
(\star)_b\Big]\Big\rangle =0.
\end{eqnarray}

\noindent Summing all these contributions with the one-loop
result, we obtain

\begin{eqnarray}\label{Beta_Result}
\frac{\beta(\alpha_0)}{\alpha_0^2} = 16\pi (A_1 + A_2 + A_3).
\end{eqnarray}

\noindent Thus, the general results, presented above, agree with
the explicit three-loop calculations, made by a different method.

Also it is possible to verify expression
(\ref{Total_Derivative_Final_Result}). After a calculation of
Feynman diagrams we obtained

\begin{eqnarray}
&& \mbox{tr}\Big\langle\ln (\star)\Big\rangle = \Bigg\{-\int
\frac{d^4k}{(2\pi)^4} \frac{d^4q}{(2\pi)^4} \frac{e^2}{k^2 R_k^2
(q^2+M^2)((q+k)^2+M^2)} \Bigg[ R_k
\Big(1+\frac{e^2}{4\pi^2}\ln\frac{\Lambda}{\mu}\Big)\nonumber\\
&& - 2 e^2 \Bigg(\int \frac{d^4t}{(2\pi)^4}\,\frac{1}{t^2 (k+t)^2}
- \sum\limits_J c_J \int \frac{d^4t}{(2\pi)^4}\,
\frac{1}{(t^2+M_J^2) ((k+t)^2+M_J^2)} \Bigg)\Bigg]\nonumber\\
&& + \int \frac{d^4q}{(2\pi)^4} \frac{d^4k}{(2\pi)^4}
\frac{d^4l}{(2\pi)^4} \frac{e^4}{k^2 R_k l^2 R_l} \Bigg(-
\frac{q^2-M^2}{(q^2+M^2)^2
((q+k)^2+M^2) ((q+l)^2+M^2)}\nonumber\\
&& - \frac{(q+k)^2-M^2}{(q^2+M^2) ((q+k)^2+M^2)^2 ((q+l+k)^2+M^2)}
+ \frac{4}{3 (q^2+M^2) ((q+k)^2+M^2)}
\nonumber\\
&& \times \frac{1}{((q+l)^2+M^2)} + \frac{8}{3 (q^2+M^2)
((q+k)^2+M^2) ((q+k+l)^2+M^2)}
\nonumber\\
&& - \frac{(2q+k+l)^2 + 2M^2}{(q^2+M^2) ((q+k)^2+M^2)
((q+l)^2+M^2) ((q+k+l)^2+M^2)} \Bigg)\Bigg\} \nonumber\\
&& \times \frac{i(\bar D^2 D^2 + D^2\bar D^2)}{8} e^{-q_\alpha
(x-y)^\alpha} \delta^4(\theta_x-\theta_y).\vphantom{\frac{1}{2}}
\end{eqnarray}

\noindent where $\mbox{tr}$ is a usual matrix trace, which (unlike
$\mbox{Tr}$) does not contain $\int d^8x$. As a consequence, after
some simple algebra we obtain

\begin{eqnarray}
&& \frac{\beta(\alpha_0)}{\alpha_0^2} = 2\pi \frac{d}{d\ln\Lambda}
\sum\limits_{I} c_I \int \frac{d^4q}{(2\pi)^4}
\frac{\partial}{\partial q^\mu} \frac{\partial}{\partial q_\mu}
\frac{\ln(q^2+M^2)}{q^2} + 4\pi \frac{d}{d\ln\Lambda} \int
\frac{d^4q}{(2\pi)^4} \frac{d^4k}{(2\pi)^4} \frac{e^2}{k^2 R_k^2}
\nonumber\\
&& \times \frac{\partial}{\partial q^\mu} \frac{\partial}{\partial
q_\mu} \Bigg(\frac{1}{q^2 (k+q)^2} - \sum\limits_I c_I
\frac{1}{(q^2+M_I^2)((k+q)^2 + M_I^2)}\Bigg) \Bigg[ R_k
\Big(1+\frac{e^2}{4\pi^2}\ln\frac{\Lambda}{\mu}\Big) \nonumber\\
&& - 2 e^2 \Bigg(\int \frac{d^4t}{(2\pi)^4}\,\frac{1}{t^2 (k+t)^2}
-\sum\limits_J c_J \int \frac{d^4t}{(2\pi)^4} \frac{1}{(t^2+M_J^2)
((k+t)^2+M_J^2)} \Bigg)\Bigg] \nonumber\\
&& + 4\pi \frac{d}{d\ln\Lambda} \int \frac{d^4q}{(2\pi)^4}
\frac{d^4k}{(2\pi)^4} \frac{d^4l}{(2\pi)^4} \frac{e^4}{k^2 R_k l^2
R_l} \frac{\partial}{\partial q^\mu} \frac{\partial}{\partial
q_\mu}\Bigg\{\Bigg( - \frac{2 k^2}{q^2 (q+k)^2 (q+l)^2 (q+k+l)^2}
\nonumber\\
&& + \frac{2}{q^2(q+k)^2(q+l)^2}\Bigg) - \sum\limits_I c_I \Bigg(-
\frac{2(k^2 + M_I^2)}{(q^2+M_I^2) ((q+k)^2+M_I^2) ((q+l)^2+M_I^2)}
\nonumber\\
&& \frac{1}{((q+k+l)^2+M_I^2)} + \frac{2}{(q^2+M_I^2)
((q+k)^2+M_I^2)((q+l)^2+M_I^2)} - \frac{1}{(q^2+M_I^2)^2}
\nonumber\\
&& \times \frac{4M_I^2}{((q+k)^2+M_I^2) ((q+l)^2+M_I^2)}
\Bigg)\Bigg\} -\mbox{integrals of
$\delta$-singularities}.\vphantom{\frac{1}{2}}
\end{eqnarray}

\noindent Then, using the equation

\begin{equation}
\int \frac{d^4q}{(2\pi)^4} \Bigg\{\frac{\partial}{\partial q_\mu}
\Big(\frac{q_\mu}{q^4} f(q)\Big) - 2\pi^2 \delta^4(q) f(q) \Bigg\}
= 2\int \frac{d^4q}{(2\pi)^4} \frac{1}{q^2} \frac{df}{dq^2},
\end{equation}

\noindent we again obtain

\begin{equation}
\frac{\beta(\alpha_0)}{\alpha_0^2} = 16\pi (A_1 + A_2 + A_3).
\end{equation}

\section{Conclusion}
\hspace{\parindent}

In this paper we have proved that the $\beta$-function in $N=1$
supersymmetric electrodynamics, regularized by higher derivatives,
is given by integrals of total derivatives in each order of the
perturbation theory. In particular, explicit expressions for all
total derivatives are obtained. Having compared them with the
explicit three-loop calculation (made by a different method) we
obtained the complete agrement.

Factorization of integrands into total derivatives is the origin
of the exact NSVZ $\beta$-function, because one of the loop
integrals can be taken. After this the $\beta$-function in $n$-th
loop is related with the anomalous dimension in $n-1$-th loop.
This was also proved in this paper by explicit summation of
Feynman diagrams.

\bigskip
\bigskip

\noindent {\Large\bf Acknowledgements.}

\bigskip

\noindent This work was supported by RFBR grant No 08-01-00281a.
The author is very grateful to A.L.Kataev and A.A.Slavnov for 
valuable discussions.

\appendix

\section{Derivation of expression for the two-point function}
\hspace{\parindent}\label{Appendix_Gamma_2_V}

Splitting the classical action $S$ into a quadratic part and an
interaction $S_I$, it is possible to present the generating
functional $Z$ in the form

\begin{equation}
Z = \exp\Big\{iS_I\Big(\frac{1}{i}\frac{\delta}{\delta
J},\frac{1}{i}\frac{\delta}{\delta j}\Big)\Big\} \exp\Big\{ i\int
d^8x\,\Big(- J \frac{e_0^2}{\partial^2 R} J + j^*
\frac{1}{\partial^2} j + \widetilde j^* \frac{1}{\partial^2}
\widetilde j\Big)\Big\} \equiv e^{iS_I} Z_0,
\end{equation}

\noindent where the Pauli--Villars determinants are omitted for
simplicity. Differentiating this generating functional with
respect to the source $J$ we obtain

\begin{eqnarray}
&& \frac{\delta Z}{\delta J_x} =  \exp(iS_I) (-2ie_0^2)
\frac{1}{\partial_x^2 R} J_x Z_0 = -2ie_0^2 \frac{1}{\partial_x^2
R} J_x Z - 2ie_0^2 \frac{1}{\partial_x^2 R} \frac{\delta S}{\delta
V_x}\Big(\frac{1}{i}\frac{\delta}{\delta
J},\frac{1}{i}\frac{\delta}{\delta j}\Big) Z;\nonumber\\
&& \frac{\delta^2 Z}{\delta J_x \delta J_y} = \exp(iS_I) \Big(
-\frac{2ie_0^2}{\partial^2 R}\delta^8_{xy} - 4e_0^2
\frac{1}{\partial_x^2 R} J_x \frac{1}{\partial_y^2 R} J_y\Big)
Z_0.
\end{eqnarray}

\noindent As a consequence

\begin{eqnarray}
&& \frac{1}{Z} \frac{\delta Z}{\delta J_x} = - \frac{2i
e_0^2}{\partial_x^2 R} J_x - \frac{2i e_0^2}{\partial_x R}
\Big\langle \frac{\delta S_I}{\delta V_x} \Big\rangle;\nonumber\\
&& \frac{1}{Z} \frac{\delta^2 Z}{\delta J_x \delta J_y} = -
\frac{2i e_0^2}{\partial_x^2 R} \delta^8_{xy} - 4e_0^4 \Big\langle
\frac{1}{\partial_x^2 R} \frac{\delta S_I}{\delta V_x}
\frac{1}{\partial_y^2 R} \frac{\delta S_I}{\delta V_y} \Big\rangle
+ 4e_0^2i\frac{1}{\partial_x^2 R} \frac{1}{\partial_y^2
R}\Big\langle \frac{\delta^2 S_I}{\delta V_x\delta
V_y}\Big\rangle.\qquad
\end{eqnarray}

\noindent (In the last equation we set $J=0$.) Therefore, if we
denote an argument of the effective action by ${\bf V}$, the
two-point Green function of the gauge superfield will satisfy

\begin{eqnarray}
&& \Bigg(\frac{\delta^2\Gamma}{\delta {\bf V}_x \delta {\bf
V}_y}\Bigg)^{-1} = - \frac{\delta^2 W}{\delta J_x \delta J_y} = i
\frac{\delta^2\ln
Z}{\delta J_x \delta J_y} = \nonumber\\
&& = \frac{2e_0^2}{\partial_x^2 R} \delta^8_{xy}  - \Bigg(4i e_0^4
\Big\langle\frac{1}{\partial_x^2 R} \frac{\delta S_I}{\delta V_x}
\frac{1}{\partial_y^2 R} \frac{\delta S_I}{\delta V_y}\Big\rangle
+ 4e_0^4 \frac{1}{\partial_x^2 R} \frac{1}{\partial_y^2
R}\Big\langle \frac{\delta^2S_I}{\delta V_x \delta
V_y}\Big\rangle\Bigg)_{\mbox{\scriptsize connected}}.\qquad
\end{eqnarray}

\noindent The inverse matrix is evidently given by

\begin{eqnarray}
\frac{\delta^2\Gamma}{\delta {\bf V}_x \delta {\bf V}_y} =
\frac{1}{2e_0^2}
\partial^2 R \delta^8_{xy} + \Big\langle i \frac{\delta S_I}{\delta V_x}
\frac{\delta S_I}{\delta V_y} + \frac{\delta^2 S_I}{\delta V_x
\delta V_y} \Big\rangle_{\mbox{\scriptsize 1PI}},
\end{eqnarray}

\noindent where the symbol 1PI means that it is necessary to keep
only one-particle irreducible graphs in this expression. Thus, a
part of the effective action quadratic in the gauge superfield can
be written as

\begin{eqnarray}\label{Gamma_2_V_Appendix}
\Gamma^{(2)}_{\bf V} = \frac{1}{4e^2} \int d^8x\, {\bf
V}\partial^2 R {\bf V} + \frac{1}{2} \int d^8x\,d^8y\,{\bf V}_x
{\bf V}_y \Big\langle i \frac{\delta S_I}{\delta V_x} \frac{\delta
S_I}{\delta V_y} + \frac{\delta^2 S_I}{\delta V_x \delta V_y}
\Big\rangle_{\mbox{\scriptsize 1PI}}.
\end{eqnarray}

\noindent The Pauli--Villars determinants can be considered
similarly. In this case $S_I$ contains the Pauli--Villars fields,
and $\sum\limits_J c_J$ should be also included.

\section{Prove of identity (\ref{Triple_Identity})}
\hspace{\parindent}\label{Appendix_Identity}

Let us consider

\begin{eqnarray}
&& X\equiv \mbox{Tr}\Big\{\theta^4 \Big( (\gamma^\mu)^{ab}
[y_\mu^*,A] [\bar\theta_b, B\}[\theta_a, C\} + (\gamma^\mu)^{ab}
(-1)^{P_A}
[\theta_a,B\} [\bar\theta_b, C\} [y_\mu^*,A]\nonumber\\
&& -4i [\theta^a,[\theta_a,
A\}\}[\bar\theta^b,B\}[\bar\theta_b,C\}\Big)\Big\} +\mbox{cyclic
perm. of $A$, $B$, $C$},
\end{eqnarray}

\noindent where $A$, $B$, and $C$ are differential operators,
containing supersymmetric covariant derivatives. It is important
that they do not explicitly depend on $\theta$. Certainly, we
assume that

\begin{equation}
P_A + P_B + P_C = 0(\mbox{mod}\ 2).
\end{equation}

\noindent Using the identities

\begin{eqnarray}
&& \mbox{Tr} \Big([y_\mu^*,A] B\Big) = - \mbox{Tr}\Big(A[y_\mu^*,B]\Big);\nonumber\\
&& (-1)^{P_B} [[A,B\},C\} + (-1)^{P_A}[[C,A\},B\} +
(-1)^{P_C}[[B,C\},A\} =0,
\end{eqnarray}

\noindent we obtain

\begin{eqnarray}
&& X = (\gamma^\mu)^{ab} \mbox{Tr} \Big\{\theta^4
\Big(\Big[y_\mu^*, A [\bar\theta_b, B\}[\theta_a, C\} + (-1)^{P_A}
[\theta_a,B\} [\bar\theta_b, C\} A\Big]
- A [\bar\theta_b,[y_\mu^*,B]\} \nonumber\\
&& \times [\theta_a, C\} - A [\bar\theta_b, B\}[\theta_a,
[y_\mu^*, C]\} - (-1)^{P_A}[\theta_a, [y_\mu^*,B]\} [\bar\theta_b,
C\}A - (-1)^{P_A}[\theta_a,B\}
\nonumber\\
&& [\bar\theta_b, [y_\mu^*, C]\}A\Big)\Big\} -4i \mbox{Tr}
\Big(\theta^4 [\theta^a,[\theta_a,
A\}\}[\bar\theta^b,B\}[\bar\theta_b,C\}\Big) +\mbox{cyclic
perm. of $A$, $B$, $C$}.\nonumber\\
\end{eqnarray}

\noindent The similar operation is repeated for $\theta$-s in
double commutators:

\begin{eqnarray}\label{X1}
&& X = (\gamma^\mu)^{ab} \mbox{Tr} \Big\{\theta^4
\Big(\Big[y_\mu^*, A [\bar\theta_b, B\}[\theta_a, C\} + (-1)^{P_A}
[\theta_a,B\} [\bar\theta_b, C\} A\Big]+ (-1)^{P_A} [\bar\theta_b,
A\}
\nonumber\\
&& \times [y_\mu^*,B] [\theta_a, C\} + (-1)^{P_B} A [y_\mu^*,B]
[\bar\theta_b, [\theta_a, C\}\} - (-1)^{P_C} [\theta_a, A\}
[\bar\theta_b, B\} [y_\mu^*, C]
\vphantom{\Big(}\nonumber\\
&& - (-1)^{P_B} A [\theta_a, [\bar\theta_b, B\}\} [y_\mu^*, C] -
[y_\mu^*,B] [\bar\theta_b, C\}[\theta_a, A\} + (-1)^{P_C}
[y_\mu^*,B] [\theta_a,[\bar\theta_b, C\}\}A
\vphantom{\Big(}\nonumber\\
&& - (-1)^{P_C}[\bar\theta_b,[\theta_a,B\}\} [y_\mu^*, C]A +
(-1)^{P_B}[\theta_a,B\} [y_\mu^*, C] [\bar\theta_b, A\} \Big)\Big)
\vphantom{\Big(}\nonumber\\
&& -4i \mbox{Tr} \Big(\theta^4 [\theta^a,[\theta_a,
A\}\}[\bar\theta^b,B\}[\bar\theta_b,C\}\Big\} +\mbox{cyclic perm.
of $A$, $B$, $C$}.
\end{eqnarray}

\noindent In addition to $\theta^4$, the commutators with
$y_\mu^*$ give one more degree of $\theta$:

\begin{equation}
[y_\mu^*,A] = -2i (\gamma^\mu)^{ab}\theta_a [\bar\theta_b,A\} +
O(\theta^0).
\end{equation}

\noindent That is why it is necessary to be careful commuting
$\theta^4$ with $A$, $B$ and $C$. For example, taking into account
that all expressions containing $\theta$ in less than fourth power
vanish, we obtain

\begin{eqnarray}
&& (-1)^{P_A} (\gamma^\mu)^{ab}\mbox{Tr}\,\theta^4 [\bar\theta_b,
A\} [y_\mu^*,B] [\theta_a, C\} =
(-1)^{P_B+1}(\gamma^\mu)^{ab}\mbox{Tr}\Big(\theta^4 [\theta_a, C\}
[\bar\theta_b, A\} [y_\mu^*,B]
\quad\nonumber\\
&& - 2\bar\theta^c\bar\theta_c \theta^d [\theta_d,[\theta_a,C\}\}
[\bar\theta_b, A\}
(-2i)(\gamma_\mu)^{ef} \theta_e [\bar\theta_f,B\}\Big)\qquad\nonumber\\
&& = \mbox{Tr}\Big((-1)^{P_B+1} \theta^4 (\gamma^\mu)^{ab}
[\theta_a, C\} [\bar\theta_b, A\} [y_\mu^*,B] + 4i \theta^4
[\theta^a,[\theta_a,C\}\} [\bar\theta^b, A\}
[\bar\theta_b,B\}\Big).
\end{eqnarray}

\noindent Similarly one can derive the following identities:

\begin{eqnarray}
&& (-1)^{P_B} (\gamma^\mu)^{ab}
\mbox{Tr}\Big(\theta^4[\theta_a,B\}[y_\mu^*,C][\bar\theta_b,A\}\Big)
\\
&& =
\mbox{Tr}\Big(-\theta^4(\gamma^\mu)^{ab}[y_\mu^*,C][\bar\theta_b,A\}[\theta_a,B\}
+4i\theta^4 [\theta^a,[\theta_a,B\}\}[\bar\theta^b,C\}[\bar\theta_b,A\}\Big);\nonumber\\
&& (-1)^{P_B+1} (\gamma^\mu)^{ab} \mbox{Tr}\Big(\theta^4 A
[\theta_a,[\bar\theta_b,B\}\}[y_\mu^*,C]\Big)
\\
&& =\mbox{Tr}\Big((-1)^{P_C+1}\theta^4
(\gamma^\mu)^{ab}[\theta_a,[\bar\theta_b,B\}\}[y_\mu^*,C]A
-4i (-1)^{P_A}\theta^4 [\theta^a, A\}[\theta_a,
[\bar\theta^b, B\}\}[\bar\theta_b,C\}\Big);\nonumber\\
&& (-1)^{P_B} (\gamma^\mu)^{ab} \mbox{Tr}\Big(\theta^4 A
[y_\mu^*,B][\bar\theta_b,[\theta_a, C\}\}\Big)
\\
&& = (-1)^{P_C}\mbox{Tr}\Big(\theta^4 (\gamma^\mu)^{ab}
[y_\mu^*,B][\bar\theta_b, [\theta_a,C\}\}A + 4i \theta^4
[\theta^a, A\}[\bar\theta_b, B\} [\bar\theta^b, [\theta_a,
C\}\}\Big).\nonumber
\end{eqnarray}

\noindent Using these equations, after some algebraic
transformations $X$ can be rewritten as

\begin{eqnarray}
&& X = \mbox{Tr}\Big\{\theta^4 (\gamma^\mu)^{ab}
\Big(\Big[y_\mu^*, A[\bar\theta_b,B\}[\theta_a,C\} +
(-1)^{P_A}[\theta_a,B\}[\bar\theta_b,C\}A \Big] -2 [y_\mu^*, A]
[\bar\theta_b, B\}
\nonumber\\
&& \times [\theta_a, C\} -2 (-1)^{P_A} [\theta_a,
B\}[\bar\theta_b, C\}[y_\mu^*,A]\Big) + 8i \theta^4
[\theta^a,[\theta_a,A\}\} [\bar\theta^b,B\}[\bar\theta_b,C\}
\Big\}
\nonumber\\
&& +\mbox{cyclic perm. of $A$, $B$, $C$}.
\end{eqnarray}

\noindent Comparing this expression with the definition of $X$, we
obtain

\begin{eqnarray}
&& X = -2X + \Bigg(\mbox{Tr}\Big(\theta^4 (\gamma_\mu)^{ab}
\Big[y_\mu^*, A [\bar\theta_b,B\} [\theta_a, C\} + (-1)^{P_A}
[\theta_a,B\}[\bar\theta_b, C\}A \Big]\Big)\nonumber\\
&&+\mbox{cyclic perm. of $A$, $B$, $C$}\smash{\Bigg)}.
\end{eqnarray}

\noindent Therefore,

\begin{eqnarray}
&& X = \frac{1}{3}\mbox{Tr}\Big(\theta^4 (\gamma_\mu)^{ab}
\Big[y_\mu^*, A [\bar\theta_b,B\} [\theta_a, C\} + (-1)^{P_A}
[\theta_a,B\}[\bar\theta_b, C\}A \Big]
\nonumber\\
&&+\mbox{cyclic perm. of $A$, $B$, $C$}.
\end{eqnarray}

\noindent This completes the proof.

\section{Simplification of expression
(\ref{Second_Sum_Total_Derivative2})}
\hspace{\parindent}\label{Appendix_Coefficient}

Let us calculate

\begin{eqnarray}\label{A_Singularities}
&& - \frac{2(\gamma^\mu)^{cd}}{n(n+1)(n+2)}\mbox{Tr}\Big\langle
\theta^4  \Big[ y_\mu^*, *^4 (\bar I_1)_d * (I_1)_c + \Big(*^3
(I_1)_c * + *^2 (I_1)_c *^2
+ *(I_1)_c *^3\Big)*\nonumber\\
&& \times (\bar I_1)_d* I_0 + *(\bar I_1)_d \Big(*^3 (I_1)_c * +
*^2 (I_1)_c *^2 + *(I_1)_c *^3\Big) + * (I_1)_c * (\bar I_1)_d *^3
\nonumber\\
&& + I_0\Big(*^3 (\bar I_1)_d * + *^2 (\bar I_1)_d *^2 + *(\bar
I_1)_d *^3\Big)*(I_1)_c * + (I_1)_c \Big(*^3 (\bar I_1)_d * + *^2
(\bar I_1)_d *^2
\nonumber\\
&& + *(\bar I_1)_d *^3\Big)* + *^2 (\bar I_1)_d *^2 (I_1)_c* +
*(I_1)_c *^2 (\bar I_1)_d*^2 \Big]\Big\rangle_n
\end{eqnarray}

\noindent for a diagram containing $n$ vertexes on the matter loop
to that external lines are attached. Due to a possibility of
making cyclic permutations (\ref{Cycle_Permutation})

\begin{equation}
(\gamma^\mu)^{cd} \mbox{Tr}\Big\langle\Big[y_\mu^*, (I_1)_c (*)_a
(\bar I_1)_d (*)_b \Big]\Big\rangle_n = -(\gamma^\mu)^{cd}
\mbox{Tr} \Big\langle\Big[y_\mu^*,\Big\langle (\bar I_1)_d (*)_b
(I_1)_c (*)_a \Big]\Big\rangle_n.\qquad
\end{equation}

\noindent Therefore, using the identity

\begin{equation}
* I_0 * = - * + *^2
\end{equation}

\noindent and Eq. (\ref{Star_Coefficients}), the considered
expression can be written as

\begin{equation}
\sum\limits_{a+b+2=n} c_a (\gamma^\mu)^{cd} \mbox{Tr}\Big\langle
\Big[y_\mu^*, (I_1)_c (*)_a (\bar I_1)_d (*)_b \Big]
\Big\rangle_n.
\end{equation}

\noindent (Two vertexes correspond to $(I_1)_c$ and $(\bar
I_1)_d$.) In order to find the coefficients $c_a$, it is necessary
to calculate all $(*^k)_a$ and $(*^k)_b$, using Eq.
(\ref{Star_Coefficients}). The result is proportional to

\begin{eqnarray}
&&\hspace*{-5mm} -\frac{1}{6} (a+1)(a+2)(a+3) - \frac{1}{2} (b+1)(b+2)(a+1)
- \frac{1}{2} (b+1)(a+1)(a+2)
-\frac{1}{6} (a+1)\nonumber\\
&&\hspace*{-5mm} \times (a+2)(a+3) + \frac{1}{6}(b+1)(b+2)(b+3)(a+1)
+ \frac{1}{4}(b+1)(b+2)(a+1)(a+2)
\nonumber\\
&&\hspace*{-5mm} + \frac{1}{6}(b+1)(a+1)(a+2)(a+3)
- \frac{1}{2} (a+1)(b+1)(b+2)
- \frac{1}{2} (a+1)(a+2)(b+1)
\nonumber\\
&&\hspace*{-5mm} -\frac{1}{6}(a+1)(a+2)(a+3) + \frac{1}{6} (b+1)(b+2)(b+3) +
\frac{1}{2}(a+1)(a+2)(b+1)
\nonumber\\
&&\hspace*{-5mm} +\frac{1}{2} (a+1)(b+1)(b+2) + \frac{1}{6} (b+1)(b+2)(b+3) -
\frac{1}{6}(b+1)(a+1)(a+2)(a+3)
\nonumber\\
&&\hspace*{-5mm} -\frac{1}{4}(a+1)(a+2)(b+1)(b+2) -
\frac{1}{6}(a+1)(b+1)(b+2)(b+3) + \frac{1}{2}(a+1)(a+2)(b+1)
\nonumber\\
&&\hspace*{-5mm} +\frac{1}{2}(a+1)(b+1)(b+2) +\frac{1}{6}(b+1)(b+2)(b+3)
-\frac{1}{2} (a+1)(a+2)(b+1)\nonumber\\
&& + \frac{1}{2} (a+1)(b+1)(b+2)
\nonumber\\
&&\hspace*{-5mm} = -\frac{1}{2} (a+1)(a+2)(a+3) - \frac{1}{2}(a+1)(a+2)(b+1)
+ \frac{1}{2} (a+1)(b+1)(b+2)
\nonumber\\
&&\hspace*{-5mm}
+ \frac{1}{2}(b+1)(b+2)(b+3)
= \frac{1}{2}(b-a)(n+1)(n+2),
\end{eqnarray}

\noindent where the sequence of terms in the first expression
corresponds to the one in Eq. (\ref{A_Singularities}). Therefore,
expression (\ref{A_Singularities}) can be rewritten as

\begin{equation}
-\sum\limits_{a+b+2=n} \frac{b-a}{n} (\gamma^\mu)^{cd} \mbox{Tr}\Big\langle
\Big[y_\mu^*, (I_1)_c (*)_a (\bar I_1)_d (*)_b \Big]\Big\rangle.
\end{equation}

\noindent Adding to this expression the first term in Eq.
(\ref{Second_Sum_Total_Derivative2}), we obtain that the total
derivative in Eq. (\ref{Second_Sum_Total_Derivative2}) can be
rewritten in the following form:

\begin{equation}
-\sum\limits_{a+b+2=n} \frac{2(b+1)}{n} (\gamma^\mu)^{cd}
\mbox{Tr}\Big\langle
\Big[y_\mu^*, (I_1)_c (*)_a (\bar I_1)_d (*)_b \Big]\Big\rangle.
\end{equation}

\section{Summation of subdiagrams with two external lines}
\hspace{\parindent}\label{Appendix_Subdiagrams}

Here we describe, how subdiagrams with two external lines are
split into groups, convenient for the calculation, and present
results of this calculation. (Expressions for the left vertexes
are omitted for simplicity.)

1. Subdiagrams with a chiral left end and an antichiral right end:

\vspace*{3cm}

\begin{picture}(0,0)
\put(0,1){\includegraphics[scale=0.35]{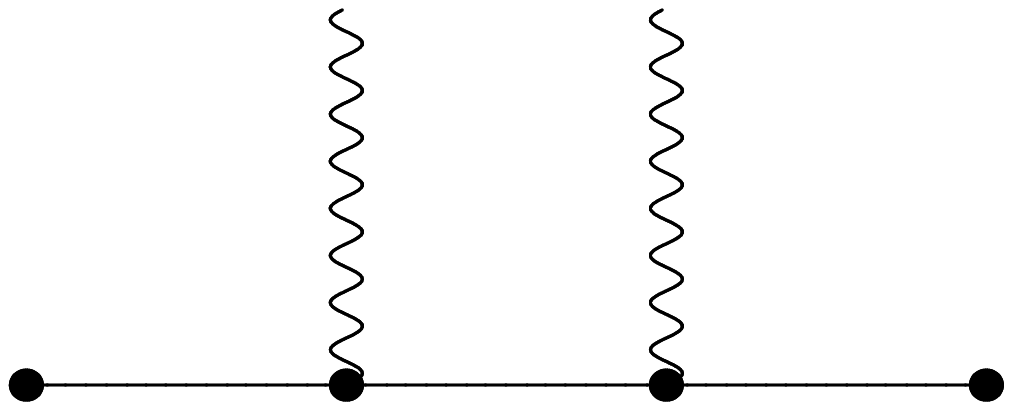}}
\put(0.8,1.3){$|$} \put(1.9,1.3){$|$} \put(3.07,1.3){$|$}
\put(5,1){\includegraphics[scale=0.35]{ver20.eps}}
\put(5.8,1.3){$|$} \put(6.92,1.3){$|$} \put(7.60,1.3){$|$}
\put(10,1){\includegraphics[scale=0.35]{ver20.eps}}
\put(12.58,1.3){$|$} \put(10.79,1.3){$|$} \put(11.42,1.3){$|$}
\put(0,-1){\includegraphics[scale=0.35]{ver20.eps}}
\put(0.78,-0.7){$|$} \put(1.42,-0.7){$|$} \put(3.05,-0.7){$|$}
\put(5.7,-1){\includegraphics[scale=0.35]{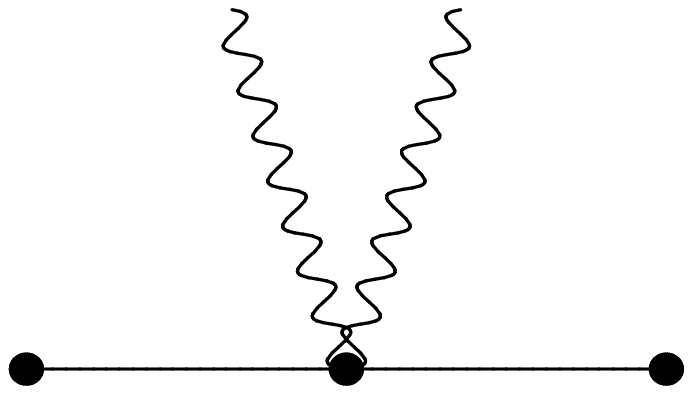}}
\put(6.35,-0.72){$|$} \put(7.5,-0.72){$|$}
\put(10.7,-1){\includegraphics[scale=0.35]{ver24.eps}}
\put(11.35,-0.72){$|$} \put(12.0,-0.72){$|$}
\put(0.5,-3){\includegraphics[scale=0.35]{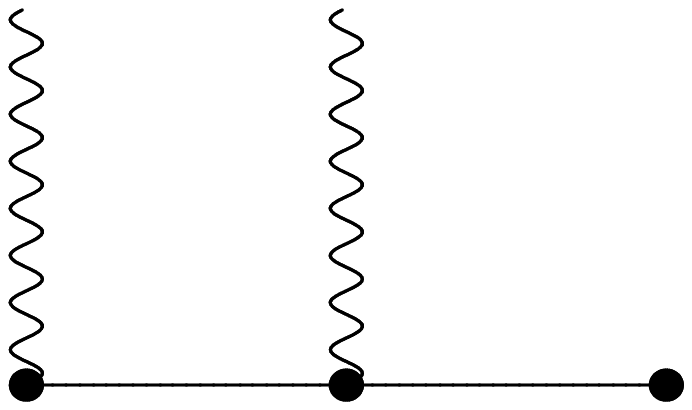}}
\put(1.16,-2.72){$|$} \put(2.3,-2.72){$|$}
\put(5.7,-3){\includegraphics[scale=0.35]{ver22.eps}}
\put(6.38,-2.72){$|$} \put(7.02,-2.72){$|$}
\put(10.7,-3){\includegraphics[scale=0.35]{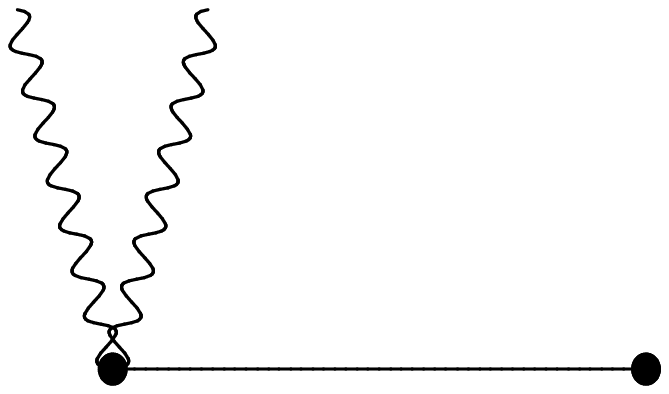}}
\put(11.7,-2.71){$|$}
\end{picture}

\vspace{3.0cm}

\noindent After the substitution ${\bf V}\to \theta^4$ the sum of these
diagrams is written as

\begin{equation}
\theta^4 \frac{M^2\bar D^2 D^2}{8 (\partial^2 + M^2)^3} = \theta^4
\frac{2 M^2}{(\partial^2 + M^2)^2}\cdot \frac{\bar D^2 D^2}{16
(\partial^2 + M^2)}.
\end{equation}

\noindent In particular, this means that in the massless case a sum
of such diagrams is equal to 0.

2. Subdiagrams with two chiral ends:

\vspace*{3cm}

\begin{picture}(0,0)
\put(0,1){\includegraphics[scale=0.35]{ver20.eps}}
\put(0.8,1.3){$|$} \put(1.9,1.3){$|$} \put(3.07,1.3){$|$}
\put(3.7,1.3){$|$}
\put(5,1){\includegraphics[scale=0.35]{ver20.eps}}
\put(5.8,1.3){$|$} \put(6.92,1.3){$|$} \put(7.60,1.3){$|$}
\put(8.7,1.3){$|$}
\put(10,1){\includegraphics[scale=0.35]{ver20.eps}}
\put(12.58,1.3){$|$} \put(10.79,1.3){$|$} \put(11.42,1.3){$|$}
\put(13.7,1.3){$|$}
\put(0,-1){\includegraphics[scale=0.35]{ver20.eps}}
\put(0.78,-0.7){$|$} \put(1.42,-0.7){$|$} \put(3.05,-0.7){$|$}
\put(3.7,-0.7){$|$}
\put(5.7,-1){\includegraphics[scale=0.35]{ver24.eps}}
\put(6.35,-0.72){$|$} \put(7.5,-0.72){$|$} \put(8.19,-0.72){$|$}
\put(10.7,-1){\includegraphics[scale=0.35]{ver24.eps}}
\put(11.35,-0.72){$|$} \put(12.0,-0.72){$|$}
\put(13.19,-0.72){$|$}
\put(0.5,-3){\includegraphics[scale=0.35]{ver22.eps}}
\put(1.16,-2.72){$|$} \put(2.3,-2.72){$|$} \put(3.0,-2.72){$|$}
\put(5.7,-3){\includegraphics[scale=0.35]{ver22.eps}}
\put(6.38,-2.72){$|$} \put(7.02,-2.72){$|$} \put(8.19,-2.72){$|$}
\put(10.7,-3){\includegraphics[scale=0.35]{ver25.eps}}
\put(11.7,-2.71){$|$}\put(13.1,-2.71){$|$}
\put(-0.3,-5){\includegraphics[scale=0.35]{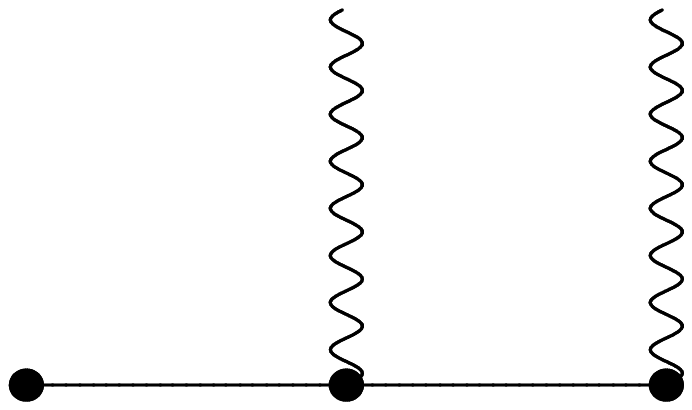}}
\put(1.08,-4.7){$|$} \put(2.2,-4.7){$|$} \put(0.44,-4.7){$|$}
\put(3.8,-5){\includegraphics[scale=0.35]{ver21.eps}}
\put(5.7,-4.7){$|$} \put(6.35,-4.7){$|$} \put(4.55,-4.7){$|$}
\put(7.8,-5){\includegraphics[scale=0.35]{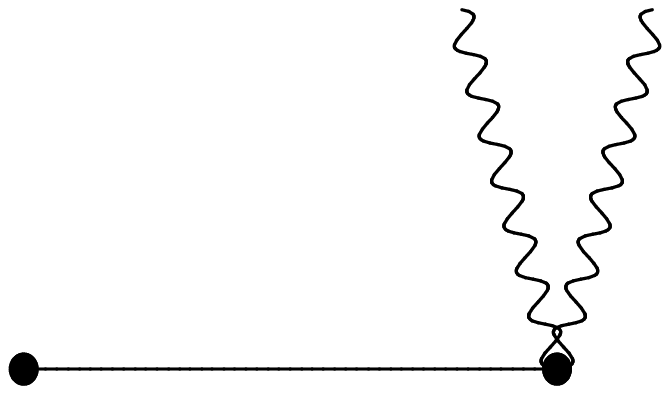}}
\put(9.8,-4.7){$|$} \put(8.48,-4.7){$|$}
\put(11.5,-5){\includegraphics[scale=0.35]{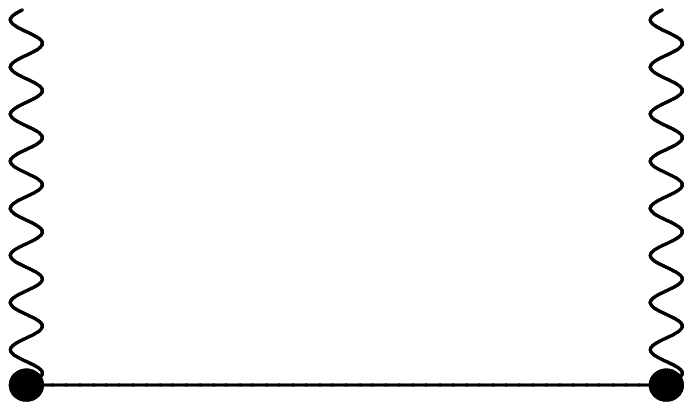}}
\put(14.1,-4.7){$|$} \put(12.32,-4.7){$|$}
\end{picture}

\vspace*{5.5cm}

\noindent These diagrams are given by

\begin{equation}
\theta^4 \frac{M^3\bar D^2}{2(\partial^2 + M^2)^3} = \theta^4
\frac{2M^2}{(\partial^2+M^2)^2}\cdot \frac{M \bar
D^2}{4(\partial^2+M^2)}.
\end{equation}

3. Subdiagrams with two antichiral ends:

\vspace*{3cm}

\begin{picture}(0,0)
\put(0,1){\includegraphics[scale=0.35]{ver20.eps}}
\put(1.4,1.3){$|$} \put(2.6,1.3){$|$}
\put(5,1){\includegraphics[scale=0.35]{ver20.eps}}
\put(6.4,1.3){$|$} \put(8.05,1.3){$|$}
\put(10,1){\includegraphics[scale=0.35]{ver20.eps}}
\put(11.9,1.3){$|$} \put(13.1,1.3){$|$}
\put(0,-1){\includegraphics[scale=0.35]{ver20.eps}}
\put(1.88,-0.7){$|$} \put(2.6,-0.7){$|$}
\put(5.7,-1){\includegraphics[scale=0.35]{ver24.eps}}
\put(7.0,-0.72){$|$}
\put(10.7,-1){\includegraphics[scale=0.35]{ver24.eps}}
\put(12.55,-0.72){$|$}
\end{picture}

\vspace*{1.5cm}

\noindent These diagrams are given by

\begin{equation}
\theta^4 \frac{D^2 M (-\partial^2)}{2(\partial^2 + M^2)^3} =
\theta^4 \Bigg(\frac{2M^2}{(\partial^2+M^2)^2}\cdot \frac{M
D^2}{4(\partial^2 + M^2)} - \frac{M D^2}{2(\partial^2 +
M^2)^2}\Bigg).
\end{equation}

4. Subdiagrams with an antichiral left end and a chiral right end:

\vspace*{3cm}

\begin{picture}(0,0)
\put(0,1){\includegraphics[scale=0.35]{ver20.eps}}
\put(1.4,1.3){$|$} \put(2.6,1.3){$|$} \put(3.7,1.3){$|$}
\put(5,1){\includegraphics[scale=0.35]{ver20.eps}}
\put(6.4,1.3){$|$} \put(8.05,1.3){$|$} \put(8.73,1.3){$|$}
\put(10,1){\includegraphics[scale=0.35]{ver20.eps}}
\put(11.9,1.3){$|$} \put(13.1,1.3){$|$} \put(13.73,1.3){$|$}
\put(0,-1){\includegraphics[scale=0.35]{ver20.eps}}
\put(1.88,-0.7){$|$} \put(2.6,-0.7){$|$} \put(3.7,-0.7){$|$}
\put(5.7,-1){\includegraphics[scale=0.35]{ver24.eps}}
\put(7.0,-0.72){$|$} \put(8.16,-0.72){$|$}
\put(10.7,-1){\includegraphics[scale=0.35]{ver24.eps}}
\put(12.55,-0.72){$|$} \put(13.2,-0.72){$|$}
\put(0.5,-3){\includegraphics[scale=0.35]{ver21.eps}}
\put(1.88,-2.7){$|$} \put(3.0,-2.7){$|$}
\put(5.7,-3){\includegraphics[scale=0.35]{ver21.eps}}
\put(7.6,-2.7){$|$} \put(8.25,-2.7){$|$}
\put(10.7,-3){\includegraphics[scale=0.35]{ver26.eps}}
\put(12.7,-2.7){$|$}
\end{picture}

\vspace*{3.0cm}

\noindent These diagrams are given by

\begin{equation}
\theta^4 \frac{2M^2 D^2\bar D^2}{(\partial^2 + M^2)^3} =
\theta^4\frac{2M^2}{(\partial^2 + M^2)}\cdot \frac{D^2 \bar
D^2}{16(\partial^2+M^2)}.
\end{equation}


\end{document}